\documentclass[aps,prb,superscriptaddress,reprint,amsmath,amssymb]{revtex4-1}

\usepackage{graphicx}
\usepackage{xcolor}

\begin{document}

\title{Phonon-limited carrier mobility and resistivity from carbon nanotubes to graphene}

\author{Jing Li}
\affiliation{Univ. Grenoble Alpes, INAC-SP2M, L\_Sim, Grenoble, France and CEA, INAC-SP2M, L\_Sim, Grenoble, France}
\author{Henrique Pereira Coutada Miranda}
\affiliation{Physics and Materials Science Research Unit, University of Luxembourg, 162a avenue de la Fa\"iencerie, L-1511 Luxembourg, Luxembourg}
\author{Yann-Michel Niquet}
\email{yniquet@cea.fr}
\affiliation{Univ. Grenoble Alpes, INAC-SP2M, L\_Sim, Grenoble, France and CEA, INAC-SP2M, L\_Sim, Grenoble, France}
\author{Luigi Genovese}
\affiliation{Univ. Grenoble Alpes, INAC-SP2M, L\_Sim, Grenoble, France and CEA, INAC-SP2M, L\_Sim, Grenoble, France}
\author{Ivan Duchemin}
\affiliation{Univ. Grenoble Alpes, INAC-SP2M, L\_Sim, Grenoble, France and CEA, INAC-SP2M, L\_Sim, Grenoble, France}
\author{Ludger Wirtz}
\affiliation{Physics and Materials Science Research Unit, University of Luxembourg, 162a avenue de la Fa\"iencerie, L-1511 Luxembourg, Luxembourg}
\author{Christophe Delerue}
\affiliation{IEMN - Dept. ISEN, UMR CNRS 8520, Lille, France}
\email{christophe.delerue@isen.fr}

\begin{abstract}
Under which conditions do the electrical transport properties of one-dimensional (1D) carbon nanotubes (CNTs) and 2D graphene become equivalent? We have performed atomistic calculations of the phonon-limited electrical mobility in graphene and in a wide range of CNTs of different types to address this issue. The theoretical study is based on a tight-binding method and a force-constant model from which all possible electron-phonon couplings are computed. The electrical resistivity of graphene is found in very good agreement with experiments performed at high carrier density. A common methodology is applied to study the transition from 1D to 2D by considering CNTs with diameter up to 16~nm. It is found that the mobility in CNTs of increasing diameter converges to the same value, the mobility in graphene. This convergence is much faster at high temperature and high carrier density. For small-diameter CNTs, the mobility strongly depends on chirality, diameter, and existence of a bandgap.
\end{abstract}

\maketitle

\section{Introduction}

Graphene and carbon nanotubes (CNTs) are respectively two-dimensional (2D) and one-dimensional (1D) allotropes of pure $sp^2$ carbon. Their remarkable physical and chemical properties have been studied extensively\cite{Saito98,Charlier07,Jorio07,Castro09} but investigations of the electrical transport in graphene and CNTs are usually done separately. Of course, it is known that graphene is a semi-metal characterized by Dirac cones at the charge neutrality point while CNTs can either be metallic or semiconducting depending on their chirality. Carrier mobilities in graphene and CNTs may reach very high values, which make them promising for high-frequency applications. However, record values differ considerably between graphene ($> 10^5$~cm$^2$/V/s)\cite{Morozov08,Bolotin08,Castro09} and semiconducting CNTs ($> 10^3$~cm$^2$/V/s).\cite{Zhou05,Ho10,Biswas11} Furthermore, one should in principle recover the transport properties of graphene by increasing the diameter of metallic or semiconducting CNTs, at least in situations where this transport is only limited by intrinsic processes like phonon scattering. Quite suprisingly, this transition of the transport properties from 1D to 2D has never been carefully investigated even if there are numerous studies on the electron-phonon coupling in the two materials.\cite{Saito98,Charlier07,Castro09,Woods00,Suzuura02,Jiang05,Jorio07,Machon06,Ferrari07,Stauber07,Suzuura08,Hwang08,Castro10,Perebeinos10,Borysenko10,Attaccalite10,Kaasbjerg12,Park14}

In this work, we study the evolution of the phonon-limited mobility of carriers from CNTs to graphene, i.e., from 1D to 2D. We calculate the electrical mobility (or resistivity) using a common methodology, enabling a direct comparison between graphene and semiconducting/metallic CNTs of varying diameters. We focus on phonon scattering not only because it is a limiting mechanism at high carrier density and high temperature \cite{Efetov10} but also because it is intrinsic to the materials, at variance with scattering by impurities or by surface optical phonons of nearby oxides, which are dependent on the geometry, on the quality of the materials and on the nature of the environment. We also leave aside carbon nanoribbons in this study, since the possible existence of edge states does not allow for a straightforward analysis of the 1D to 2D transistion.

In the following, we present fully atomistic calculations of the phonon-limited mobility in graphene and CNTs with diameter up to 16~nm.
We show that the mobility in CNTs tends to the limit in graphene at large diameter (sometimes above 10~nm), but that the route to it strongly depends on size, chirality, and temperature. These behaviors are explained by specific features of the band structure and of the electron-phonon coupling.

\section{Methodology}

\subsection{Justification of the methodology}

A strong motivation for this study is that it sheds light on the effects of confinement and dimensionality upon the electron/phonon band structures, the electron-phonon coupling and the transport.
For example, the scattering of carriers in 1D semiconductor nanostructures such as nanowires is merely limited to back-scattering when only one electronic band is populated. This tends to reduce the scattering probability.
At the same time, confinement leads to an enhancement of the electron-phonon coupling, which is stronger in 1D than in 2D (and 3D).\cite{Luisier09,Zhang10,Neophytou11} In this context, it is particularly interesting to investigate these effects in 1D and 2D carbon allotropes since their band structures considerably differ from those of usual semiconductor nanostructures. In addition, it is important to understand how (and when) the phonon-limited carrier mobility in CNTs of increasing diameter approaches the limit in graphene as a function of temperature, carrier density, chirality, and metallic or semiconducting character.

There are already numerous theoretical studies on the phonon-limited transport in graphene \cite{Stauber07,Hwang08,Castro10,Mariani10,Perebeinos10,Kaasbjerg12,Park14} and CNTs \cite{Suzuura02,Perebeinos05,Torres06,Popov06,Kauser07,Pennington07,Charlier07,Koswatta07,Zhao09} but it is only in recent works on graphene that the couplings to all phonons have been calculated from first-principles without extracting parameters from experiments.\cite{Borysenko10,Kaasbjerg12,Park14} From these works, we learn that it is essential to consider many different phonon modes like longitudinal acoustic (LA), transverse acoustic (TA) and optical phonons to predict the temperature- and density-dependent resistivity of graphene. We deduce that it is likely the case for CNTs, especially because the $k$ selection rule is broken in the direction perpendicular to the CNT axis, in particular in small-diameter CNTs.

Therefore it is essential to consider all possible electron-phonon scattering processes to predict the phonon-limited mobility in CNTs. In order to catch the 1D-2D transition, it is also necessary to go well beyond previous calculations that were limited to small-diameter tubes. However, first-principles calculations cannot be performed on large-diameter CNTs because the computational cost increases dramatically with the number of atoms per unit cell. In this work, we present calculations combining tight-binding (TB) and force-constant models for electrons and phonons, respectively. The parameters of these models were refined on the latest first-principles calculations.\cite{Park14,Mauri14} The low-field mobility is computed taking all electron and phonon bands into account, the carriers being coupled to all possible phonons, including intra- and inter-subband scattering. We show that this approach gives results for graphene in excellent agreement with first-principles calculations\cite{Park14,Mauri14} and with experiments.\cite{Efetov10,Zou10} The same methodology is then applied to CNTs. In the following, we present results for $p$-type graphene and CNTs, i.e., for the transport of holes. Very similar results are found for electrons. For the sake of comparison, carrier densities are given as equivalent surface densities (in cm$^{-2}$). For a CNT of diameter $d$, the 1D and 2D carrier densities are related by $n_{2D} = n_{1D}/(\pi d)$.

\subsection{Tight-binding Hamiltonian}

The electronic states of graphene and CNTs are written in the basis of the $p_z$ orbitals where $z$ is the axis perpendicular to the lattice. Interactions are restricted to first-nearest-neighbors (1NNs). The Hamiltonian matrix is therefore defined by two quantities, the onsite energy $E_p$ and the 1NN hopping term $V_{pp\pi}$ that both depend on the atomic displacements induced by the phonons. The variations of $V_{pp\pi}$ are governed by a power law \cite{Harrison89}

\begin{equation}
V_{pp\pi} = V^0_{pp\pi}\left(\frac{d_0}{d}\right)^{n_{pp\pi}},
\label{eq_Harrison}
\end{equation}

\noindent in which $d$ and $d_0$ are the actual and equilibrium bond lengths between 1NN atoms, respectively.
The parameters in Eq.~(\ref{eq_Harrison}) were set based on the GW calculation data in Ref.~\onlinecite{Mauri14}, i.e. $d_0=1.42$~\AA, $V^0_{pp\pi}=3.0$~eV, (which gives $10^6$~m/s for the Fermi velocity,) and $n_{pp\pi}=3.47$, (which fits the canonical acoustic gauge field, $5.52$~eV).

The variations of the onsite term $E_p$ with atomic displacements were ignored in most previous TB studies \cite{Perebeinos05,Stauber07,Mariani10,Perebeinos10} arguing that they give little contribution to the deformation potentials. In our case, we follow the arguments of Ref.~\onlinecite{Niquet09} and introduce in $E_p$ a term proportional to the average of the relative bond length variation

\begin{equation}
E_p = \frac{2}{3}\alpha_{p}\sum_{i=1}^3\frac{d_i-d_0}{d_0}
\label{eq_onsite}
\end{equation}

\noindent where the sum is over the three 1NN atoms. The Dirac point at $d=d_0$ is taken as the origin of the electronic energies.
The parameter $\alpha_{p}$ in Eq.~\ref{eq_onsite} is set to $-4.162$ eV, and has been extracted from Density Functional Theory (DFT) calculations of the band structure of graphene for different cell sizes. These DFT calculations were performed with the BigDFT\cite{BigDFT} code, using Norm-Conserving Pseudopotentials\cite{HGH-K} and the Generalized Gradient Approximation\cite{PBE} for the exchange and correlation potential.  BigDFT makes use of a systematic real-space wavelet basis, and can handle surface-like boundary conditions,\cite{PSolver} which allows to define a common energy reference for all cell sizes and enables a direct comparison of energy levels.\cite{Note}

The changes in $V_{pp\pi}$ due to bond length variations (Eq.~\ref{eq_Harrison}) contribute to the coupling of electrons to both LA and TA phonons, whereas those in $E_p$ (Eq.~\ref{eq_onsite}) are only relevant for LA phonons. Their respective influence on the mobility was heavily debated\cite{Woods00,Perebeinos05,Stauber07,Mariani10,Perebeinos10,Suzuura02} but it is now clear that both of them must be taken into account.\cite{Park14}

In the case of CNTs, the hopping terms between 1NN $p_z$ orbitals slightly vary from one bond to another due to the curvature.\cite{Dresselhaus05,Kane97b,Charlier07} As a consequence, there is a small curvature-induced gap in "metallic" zigzag CNTs scaling as $1/d^2$. For example, we obtain gaps of 80, 45, and 29 meV in (9,0), (12,0), and (15,0) CNTs, respectively, in excellent agreement with experiments.\cite{Ouyang01} For symmetry reasons, armchair CNTs always preserve their metallic character.\cite{Dresselhaus05,Kane97b,Charlier07} The $\sigma-\pi$ hybridization induced by the curvature \cite{Blase94} is neglected in the present calculations.

\subsection{Phonons}

For the calculation of the phonons, the dynamical matrix is built by a fourth-nearest-neighbour (4NN) force-constant model as originally fitted to experimental data by Jishi et al.\cite{Jishi93} and later fitted to \textit{ab-initio} calculations and enhanced by off-diagonal force-constants for the 2NN.\cite{Wirtz04} For the current calculations, we refitted the model, taking also the off-diagonal elements of the 4NN into account. The importance of the 4NN off-diagonal elements had already been noted before in \textit{ab-initio} phonon calculations.\cite{Dubay}
While the phonon-dispersion itself can be fitted very well without the off-diagonal terms, they have an important impact on the phonon modes: they are necessary to obtain the correct amplitude for the admixture of optical components to the acoustic phonon modes at non-vanishing wave-vector. 
This admixture considerably influences the contribution of the acoustic modes to the mobility.\cite{Mauri14} For example, the ratio of the acoustic phonon deformation potential with and without optical phonon mixing is about $64.2$\% in our model, which is comparable to the one in Ref.~\onlinecite{Mauri14}. The parameters of our force-constant model and the dispersion obtained from it are given in Appendix~\ref{4NN_model}.

\subsection{Electron-phonon coupling}

The same parameters are used for the calculations of the electron and phonon band structures of CNTs and graphene -- there is no additional parameter.

Following Ref.~\onlinecite{Zhang10}, the matrix element $M_{{\bf k},b}^{{\bf k}',b'}$ for the transition of an electron from an initial state $|{\bf k},b \rangle$ to a final state $|{\bf k'},b' \rangle$ $({\bf k}'={\bf k}+{\bf q})$ after emission of a phonon $|-{\bf q},j\rangle$ (same formula for absorption) is given by

\begin{eqnarray}
&&M_{{\bf k},b}^{{\bf k}',b'} = \sum_{\alpha, i} \sqrt{\frac{\hbar}{2 N M \omega_{j}({\bf q})}} e_{\alpha i}^{(j)}({\bf q}) \nonumber \\
&&\times \sum_{\beta, \beta'} C_{\beta'}^{{\bf k}',b' *} C_{\beta}^{{\bf k},b} \sum_{m,m'} e^{i{\bf k}.{\bf R}_{m \beta}} e^{-i{\bf k}'.{\bf R}_{m' \beta'}} \nonumber \\
&&\times \frac{\partial \langle \phi({\bf r}-{\bf R}_{m'\beta'}) |H| \phi ({\bf r}-{\bf R}_{m\beta}) \rangle}{\partial R_{0 \alpha i}}
\label{eqn_matrix_element}
\end{eqnarray}

\noindent where ${\bf k}$ and ${\bf q}$ are the wave vectors of the electron and phonon, respectively, while $b$ and $j$ are the indexes of the electron band and phonon mode. $C_{\beta}^{{\bf k},b}$ is an eigenvector element of the electronic Hamiltonian $H$ at equilibrium, where $\beta$ (or $\alpha$) is the atom index in the unit cell. $\phi ({\bf r}-{\bf R}_{m\beta})$ is the $p_z$ atomic orbital centered on atom $\beta$ in the unit cell $m$. $e_{\alpha i}^{(j)}({\bf q})$ is an eigenvector element of phonon state $|{\bf q},j\rangle$ and $\omega_{j}({\bf q})$ is the corresponding eigenfrequency. $R_{0 \alpha i}$ is the $i$ component ($x$,$y$,$z$) of vector ${\bf R}_{0 \alpha}$, $N$ is the number of Wigner-Seitz unit cells, and $M$ is the mass of a carbon atom. The transition rate is given by Fermi's golden rule,

\begin{eqnarray}
&&W_{{\bf k}b,{\bf k}'b'} = \frac{2 \pi}{\hbar} \sum_{j} \left| M_{{\bf k},b}^{{\bf k}',b'} \right|^{2} \nonumber \\
&&\times \bigl \{ n_{{\bf q},j} \delta(E_{{\bf k}',b'}-E_{{\bf k},b}-\hbar \omega_{j}({\bf q})) \nonumber \\
&&+ [n_{{\bf q},j}+1] \delta(E_{{\bf k}',b'}-E_{{\bf k},b}+\hbar \omega_{j}({\bf q})) \bigr \}
\label{eqn_W}
\end{eqnarray}

\noindent where $E_{{\bf k},b}$ is the energy of $|{\bf k},b \rangle$ and $n_{{\bf q},j}$ is the equilibrium phonon occupation number (Bose-Einstein distribution).

The electron-phonon scattering rates are calculated for all electronic bands within at least $5k_BT$+200 meV of the Fermi level, and for all phonon modes.

\subsection{Mobility}

The low field mobility $\mu$ is obtained by the resolution of the Boltzmann transport equation in the stationary regime. Under the application of a constant electric field $F$, the distribution function in the state $|{\bf k},b\rangle$ is given to the first-order in $F$ by $f_{b}({\bf k}) = f^{0}(E_{{\bf k},b}) + eF g_{b}({\bf k})$ where $f^0$ is the Fermi-Dirac distribution function. In CNTs, $g_{b}({\bf k})$ is solution of the following equations:

\begin{eqnarray}
&&\sum_{b'} \int g_{b}({\bf k}) \bigl \{W_{{\bf k}b,{\bf k}'b'} [1-f^0(E_{{\bf k}',b'})] + W_{{\bf k}'b',{\bf k}b} f^0(E_{{\bf k}',b'}) \bigr \} \nonumber \\
&&- g_{b'}({\bf k}') \bigl \{W_{{\bf k}'b',{\bf k}b} [1-f^0(E_{{\bf k},b})] + W_{{\bf k}b,{\bf k}'b'} f^0(E_{{\bf k},b}) \bigr \} dk' \nonumber \\
&& = \frac{2 \pi}{L} v_{b}({\bf k}) \left( \frac{\partial f^0}{\partial E} \right)_{E_{{\bf k},b}}
\label{eqn_g}
\end{eqnarray}

\noindent where $v_{b}({\bf k}) = \hbar^{-1} \partial E_{{\bf k},b} / \partial k$ is the group velocity along the electric field and $L$ is the length of the CNT. The mobility is then given by

\begin{equation}
\mu = -e \frac{\sum_{b} \int g_{b}({\bf k}) v_{b}({\bf k}) dk}{\sum_{b} \int f^0(E_{{\bf k},b}) dk}.
\label{eqn_mu}
\end{equation}

Equations~(\ref{eqn_g}-\ref{eqn_mu}) remain valid for graphene using the substitutions $dk' \to d^{2}{\bf k}'$ and $2\pi / L \to 4\pi^{2}/S$ where $S$ is the sample surface. 

The linearized Boltzmann transport equation is solved exactly. Brillouin zone integrations are performed on a non-homogeneous 1D grid for CNTs and on an triangular mesh for graphene (e.g., 8600 $k$-points in 1D and 17124 triangles in 2D for a surface carrier density of $13.6 \times 10^{12}$~cm$^{-2}$ and $T=300$~K). 

\section{Results and discussion}

\begin{figure}
\includegraphics[width=0.9\columnwidth]{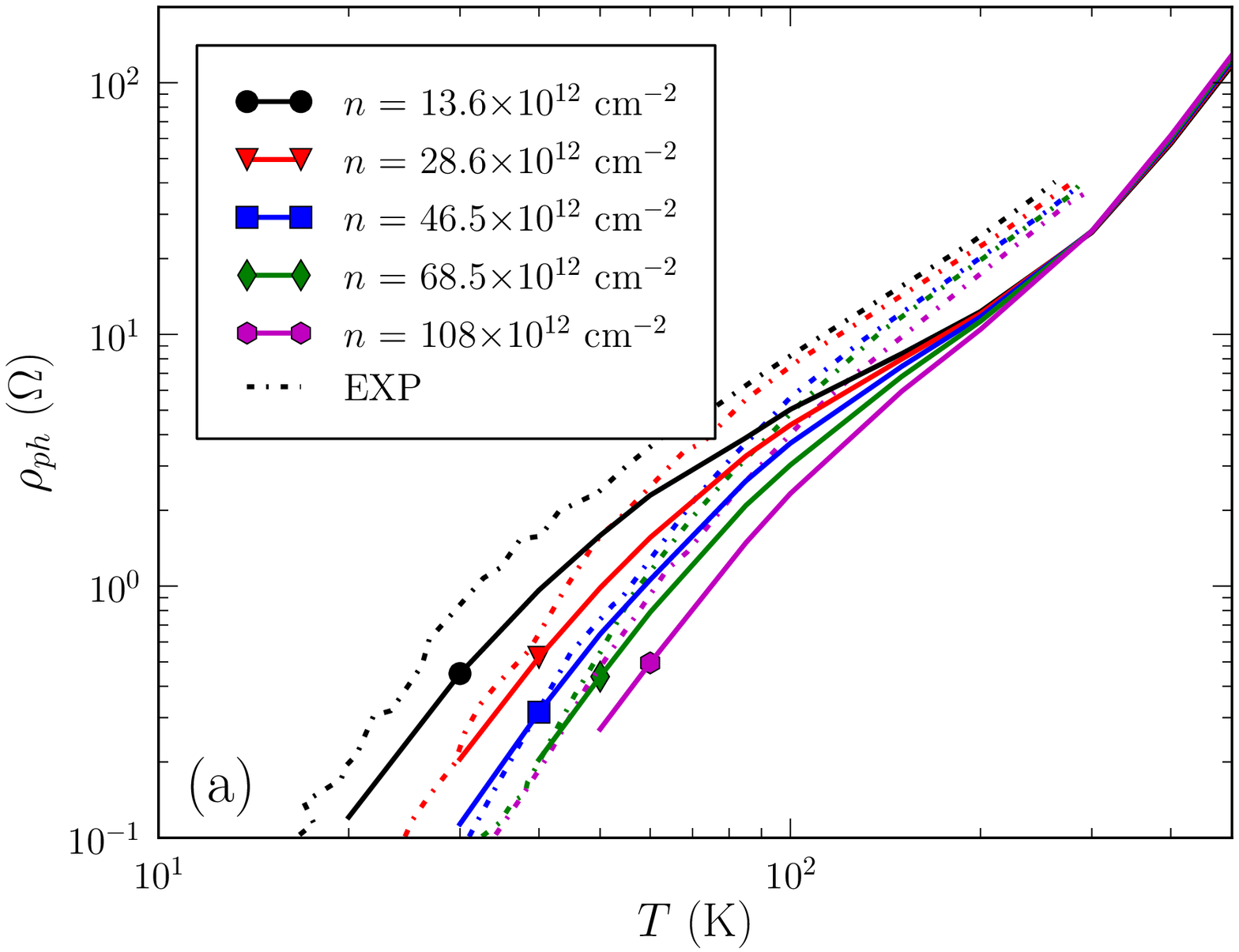}
\includegraphics[width=0.9\columnwidth]{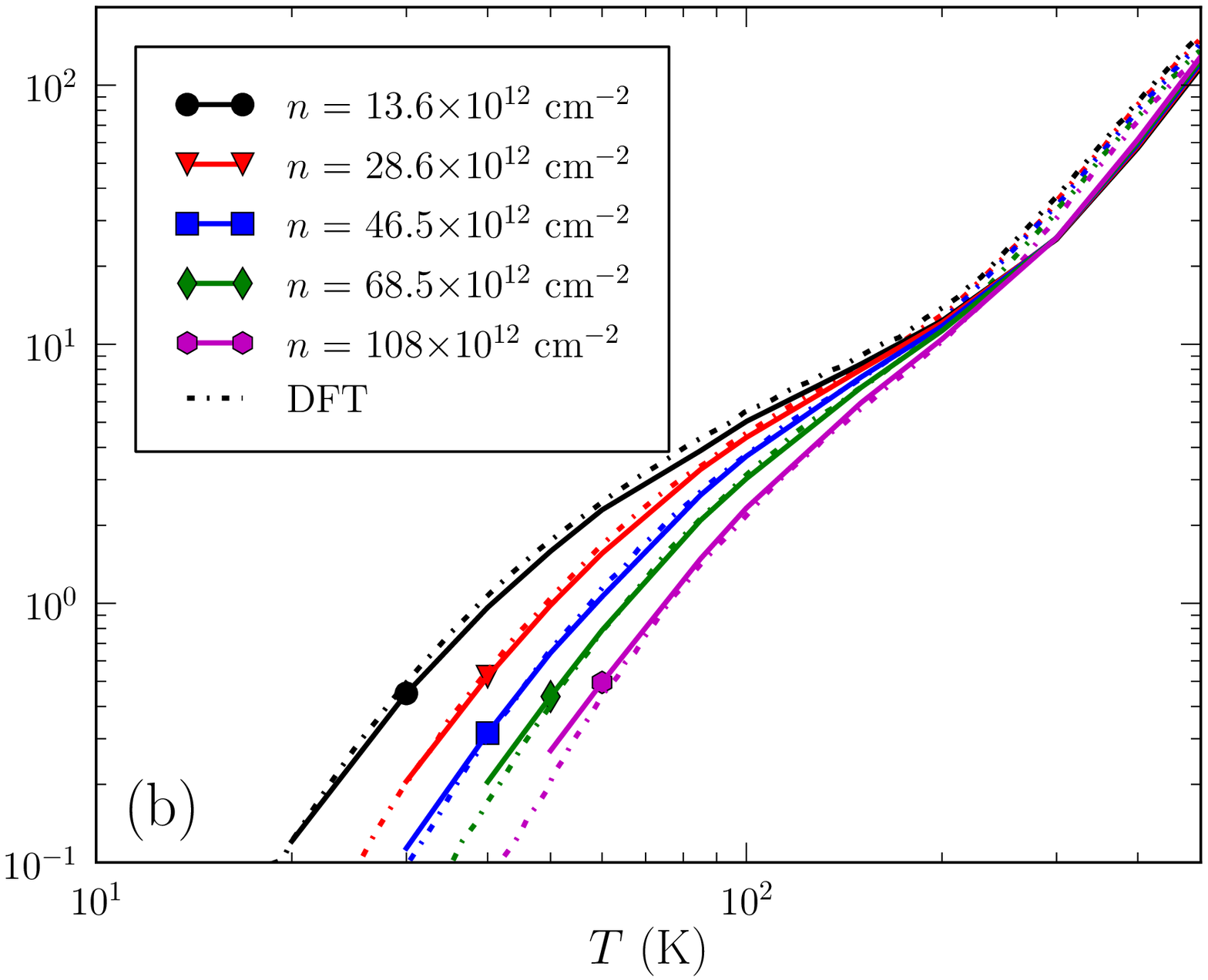}
\caption{Comparison of the resistivity of graphene versus temperature for different hole densities. (a) Present work (solid lines) and experimental data (dotted lines), which correspond to the temperature dependent part of the resistivity measured in Ref.~\onlinecite{Efetov10}. (b) Present work (solid lines) and DFT calculations.\cite{Park14}}
\label{fig_tb_dft}
\end{figure}

\subsection{Application to graphene}

Figure~\ref{fig_tb_dft} presents the electrical resistivity $\rho$ of graphene calculated versus temperature $T$, for several carrier densities $n_{2D}$. In accordance with early theoretical predictions\cite{Hwang08} and experiments, \cite{Efetov10} the resistivity is proportional to $T^4$ at low temperature, while at higher temperature it varies linearly with $T$ (the threshold strongly depends on $n_{2D}$). In the $T^4$ regime, short wavelength acoustic phonons are frozen out, restricting scattering processes to small scattering angles.\cite{Hwang08,Efetov10} In agreement with DFT calculations,\cite{Park14} both LA and TA phonons contribute to the scattering in the $T^4$ and $T$ regimes. We also confirm that: i) in the $T$ regime and above, the resistivity does not depend on $n_{2D}$;\cite{Hwang08,Efetov10} ii) at room temperature and above, the resistivity is enhanced by optical phonon scattering \cite{Borysenko10,Park14} so that it deviates from the $T$ trend. 

In Ref.~\onlinecite{Efetov10}, the resistivity of graphene was measured at high carrier density ($n_{2D} > 10^{13}$~cm$^{-3}$). Its temperature dependent component is displayed in Fig.~\ref{fig_tb_dft} and is compared to our predictions. The agreement between theory and experiments is fairly good. As shown in Appendix~\ref{resistivity_graphene}, in this range of carrier density, the resistivity is mainly limited by (intrinsic) phonon scattering, justifying the comparison. Also, the agreement is excellent with the DFT calculations of Ref.~\onlinecite{Park14} (Fig.~\ref{fig_tb_dft}b).

In Appendix~\ref{resistivity_graphene}, we show that our predicted resistivities also agree with the experimental data of Ref.~\onlinecite{Zou10} measured at lower carrier density. In that case, we have to include scattering by the surface optical (SO) phonons of the substrate. We also demonstrate in Appendix~\ref{resistivity_graphene} that scattering by intrinsic phonons becomes more efficient than scattering by SO phonons not only at low temperature because it involves lower energy phonons, but also at high temperature and high carrier density where SO phonon potentials are efficiently screened.

\begin{figure}
\includegraphics[width=0.8\columnwidth]{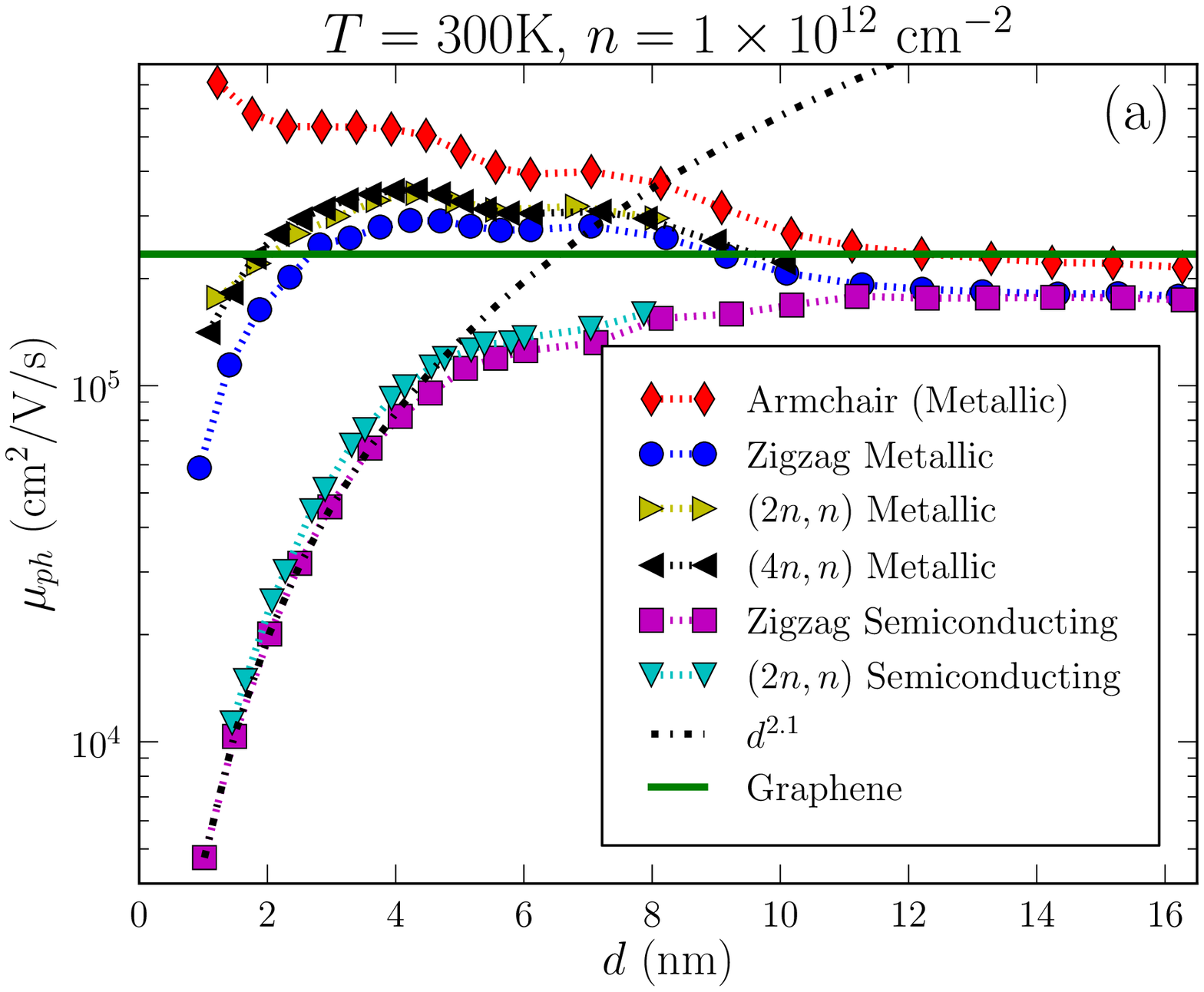}\\
\includegraphics[width=0.8\columnwidth]{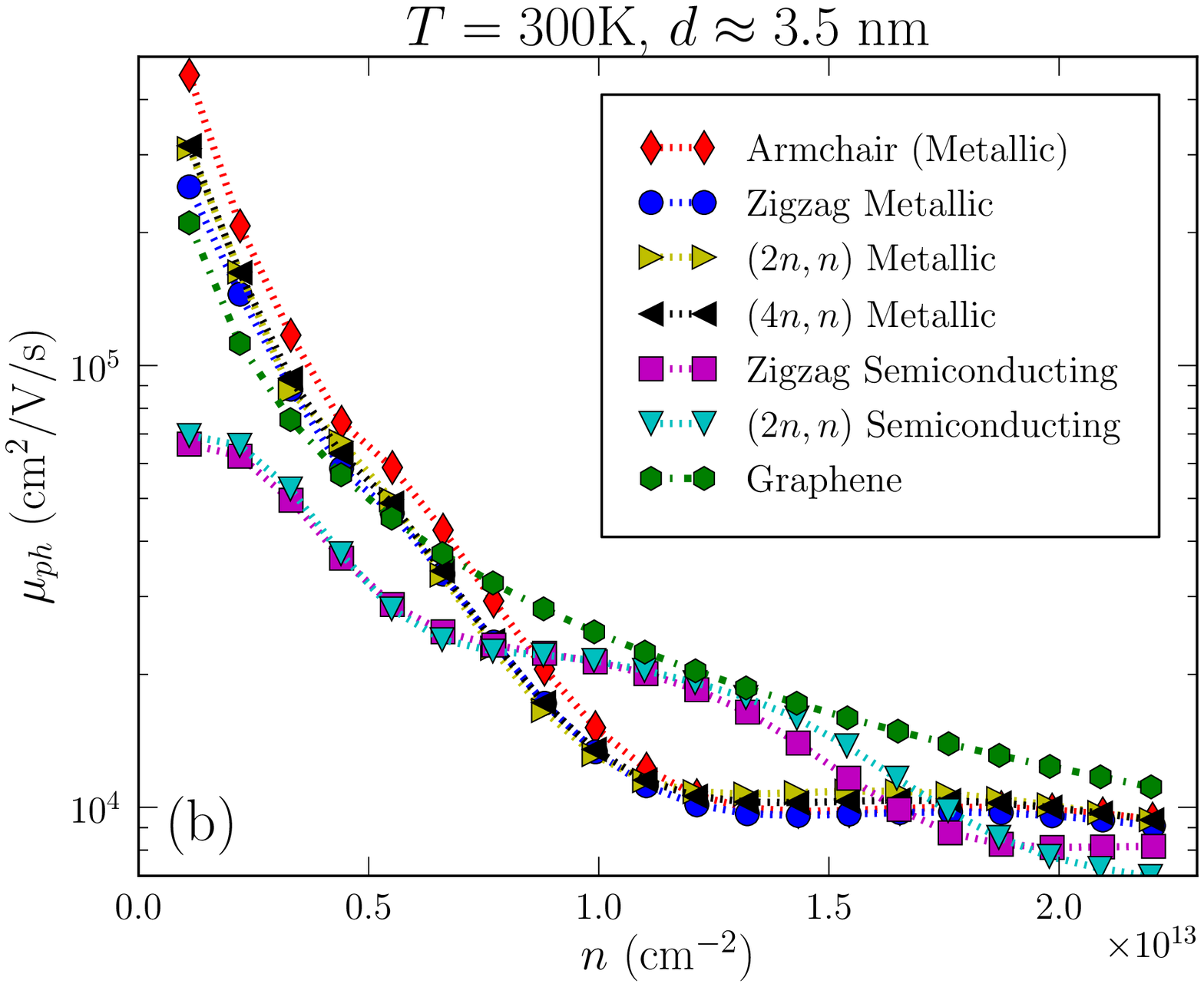}\\
\includegraphics[width=0.8\columnwidth]{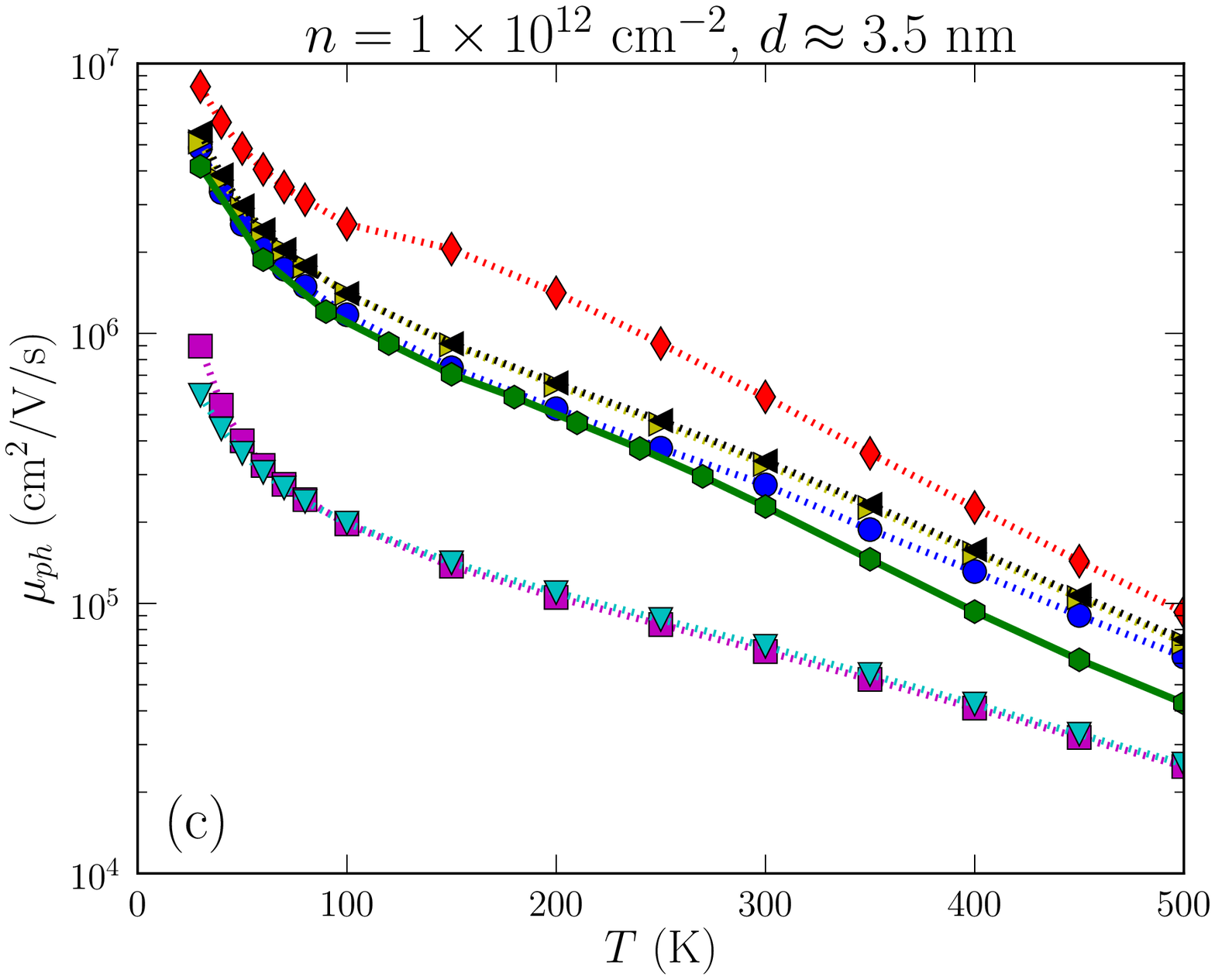}\\
\caption{Phonon-limited mobility in CNTs with different chirality (armchair $(n,n)$, zigzag $(n,0)$, chiral $(2n,n)$ and chiral $(4n,n)$) compared to graphene (green line) versus (a) diameter at 300~K and $n_{2D}=10^{12}$~cm$^{-2}$, (b) carrier density at 300~K and $d\approx3.5$~nm and, (c) temperature at $n_{2D} = 13.6 \times 10^{12}$~cm$^{-2}$ and $d\approx3.5$~nm.}
\label{fig_d_n_T}
\end{figure}

\subsection{Carrier mobility in CNTs}

We consider now zigzag $(n,0)$, armchair $(n,n)$, chiral $(2n,n)$ and chiral $(4n,n)$ CNTs. $(n,n)$ and $(4n,n)$ CNTs are always metallic. Zigzag $(n,0)$ and chiral $(2n,n)$ CNTs are metallic when $n$ is a multiple of 3, semiconducting otherwise. As discussed above, a gap is induced by the curvature in small-diameter metallic $(n,0)$ CNTs. The electronic band structures of CNTs (Appendix~\ref{band_CNT}) are intimately related to that of graphene.\cite{Saito98,Charlier07} In the 2D $k$ space of graphene, the 1D $k$ vectors allowed for CNTs form a bundle of parallel lines along the tube direction, the separation between the lines being inversely proportional to the CNT diameter. This results in CNT band structures composed of successive subbands, one for each line. In these conditions, since carrier transport takes place within an energy window of a few $k_BT$ around the Fermi surface, we anticipate that the transport properties of CNTs can reach those of graphene in three ways, namely by increasing 1) the tube diameter, 2) the carrier density, or 3) the temperature. In the three cases, the number of subbands in the transport energy window will increase in such a way that the 1D transport will progressively turn into 2D transport -- the quantification of $k$ perpendicular to the tube axis becoming irrelevant. Figure~\ref{fig_d_n_T} illustrates the evolution of the mobility along the three ways. It shows that the mobility in CNTs actually tends to its limit in graphene but with very different behaviors depending on the nature of the CNTs. The reasons why are discussed below.

\subsubsection{Mobility versus diameter}

The carrier mobility in CNTs calculated for $n_{2D}=10^{12}$~cm$^{-2}$ and $T=300$~K is plotted as a function of the diameter $d$ in Fig.~\ref{fig_d_n_T}a. The mobility is considerably smaller in semiconducting CNTs with $d<10$~nm than in metallic ones, and is also much smaller than in graphene for the same carrier density. This is mainly due to the non-zero mass of the carriers since the bands are parabolic near their extrema. For $d<4$~nm, the mobility in zigzag semiconducting CNTs scales with diameter as $d^{2.1}$, in agreement with early predictions\cite{Perebeinos05} and measurements.\cite{Zhou05} An approximately quadratic relationship is also obtained for small-diameter $(2n,n)$ semiconducting CNTs, as reported for different chiralities.\cite{Kauser07} The important reduction of the mobility in semiconducting CNTs with decreasing size is due to the increase of the effective mass \cite{Kauser07} and to the enhancement of electron-phonon coupling, in particular with radial breathing modes.\cite{Perebeinos05,Popov06,Kauser07,Charlier07} The strong enhancement of the electron-phonon coupling in small-diameter 1D conductors seems to be a general trend, as shown for example for electrons in thin Si nanowires.\cite{Zhang10,Neophytou11} For $d>10$~nm, the mobility in semiconducting zigzag CNTs saturates at a constant value, however smaller than in graphene for these particular $T$ and $n_{2D}$. Quite generally, we have found that plateaus appear in mobility curves when the Fermi level lies between two subbands and the upper subband does not contribute to transport.

For metallic CNTs, the behavior of the mobility versus diameter is very different (Fig.~\ref{fig_d_n_T}a), as expected from the presence of linearly dispersive bands (Appendix~\ref{band_CNT}).\cite{Park04,Popov06} At large diameter, the mobility is close to the value in graphene as several bands are populated, including parabolic bands that contribute to the diffusive transport. Interestingly, for $n_{2D}=10^{12}$~cm$^{-2}$ and $T=300$~K, the mobility in metallic zigzag CNTs saturates at the same plateau as in semiconducting ones for $12<d<16$~nm.

At smaller diameter ($4<d<12$~nm), only linearly dispersive bands are occupied in metallic CNTs (Appendix~\ref{band_CNT}), and the mobility increases because the number of allowed scattering processes decreases. The mobility is larger in armchair than in zigzag CNTs because there is no coupling to the LA phonons in the linear bands of armchair CNTs for symmetry reasons\cite{Popov06}. The total scattering by acoustic phonons is, therefore, much smaller in armchair CNTs (see Sect.~\ref{mu_opt} below). The difference between metallic armchair and zigzag CNTs is progressively reduced when the temperature or carrier density are increased (Fig.~\ref{fig_d_n_T}b,c), since carriers start to occupy parabolic bands in which the scattering by LA phonons is allowed, in both armchair and zigzag CNTs.\cite{Popov06}

In metallic zigzag CNTs with small diameter ($d<3$~nm), the mobility suddenly decreases because the electron-phonon interaction considerably increases, in particular with LA and radial breathing modes,\cite{Popov06} as in semiconducting CNTs. For very small diameters, the bands near the neutrality point become parabolic due to the curvature-induced gap but the effect on the mobility is small for all carrier densities considered here.

The mobility in chiral metallic CNTs [$(2n,n)$ or $(4n,n)$] follows the same trends as in zigzag metallic CNTs versus diameter, carrier density or temperature. Similarly, the $(2n,n)$ chiral CNTs show the same behavior as zigzag semiconducting CNTs. Therefore, chiral and achiral CNTs of the same nature (semiconducting/metallic) exhibit similar phonon-limited transport properties.\cite{Kauser07}

\subsubsection{Mobility versus carrier density}

We discuss now the evolution of the carrier mobility as a function of $n_{2D}$, the other parameters ($T=300$~K, $d\approx3.5$~nm) being kept constant (Fig.~\ref{fig_d_n_T}b). For carrier densities of the order of $10^{12}$~cm$^{-2}$, the mobilities of the different types of CNTs span more than one order of magnitude. Two trends can be highlighted when $n_{2D}$ is increased. First, the mobility decreases, almost exactly like $1/n_{2D}$ as in the case of graphene (see the plot of $\mu \times n_{2D}$ versus $n_{2D}$ in Appendix~\ref{mu_n}). The well-known independence of $\mu \times n_{2D}$ (or the resistivity) on the Fermi energy or equivalently on the carrier density in graphene at 300~K \cite{Hwang08,Efetov10} is approximately recovered in CNTs at high carrier density. Second, above $\sim 2 \times 10^{13}$~cm$^{-2}$, the mobilities of CNTs and graphene tend to converge to the same value. In that case, the integration over many subbands smooths out the effects resulting from the 1D character of the CNTs. For $n_{2D}$ of the order of $10^{12}$~cm$^{-2}$, the mobility is higher in metallic CNTs than in graphene but the situation is reversed when $n_{2D}$ reaches a certain threshold (for example $\sim 7 \times 10^{12}$ cm$^{-2}$ for armchair CNTs). In fact, the mobility becomes smaller in metallic CNTs than in graphene when the parabolic subbands start to be occupied (Appendix~\ref{band_CNT}). The opposite situation arises when the Fermi level only crosses the linear bands.

\subsubsection{Mobility versus temperature}

The variations of the mobility with temperature at fixed carrier density ($10^{12}$~cm$^{-2}$) and diameter (3.5~nm) are plotted in Fig.~\ref{fig_d_n_T}c. At high temperature, the mobility in CNTs decreases roughly like $1/T^{\beta}$ where the exponent $\beta$ is close to unity like in graphene but slightly differs from case to case. However, for this particular diameter and carrier density, the spread of the calculated mobilities is still important at 500~K. This shows that the broadening and the averaging induced by the temperature are not sufficient to smooth out 1D band structure effects at this diameter. As already discussed for graphene, the exponent $\beta$ differs from unity because of the increasing role of optical phonons at high temperature. However, we show in the next section that the relative importance of optical phonons at high $T$ strongly varies from one CNT to another, being huge in armchair metallic CNTs and remaining modest in semiconducting CNTs [zigzag or $(2n,n)$].

\subsubsection{Role of the different phonons}
\label{mu_opt}

In this section, we clarify the relative importance of the different phonons for each type of CNTs, and we compare with the case of graphene.

\begin{figure}
\includegraphics[width=0.9\columnwidth]{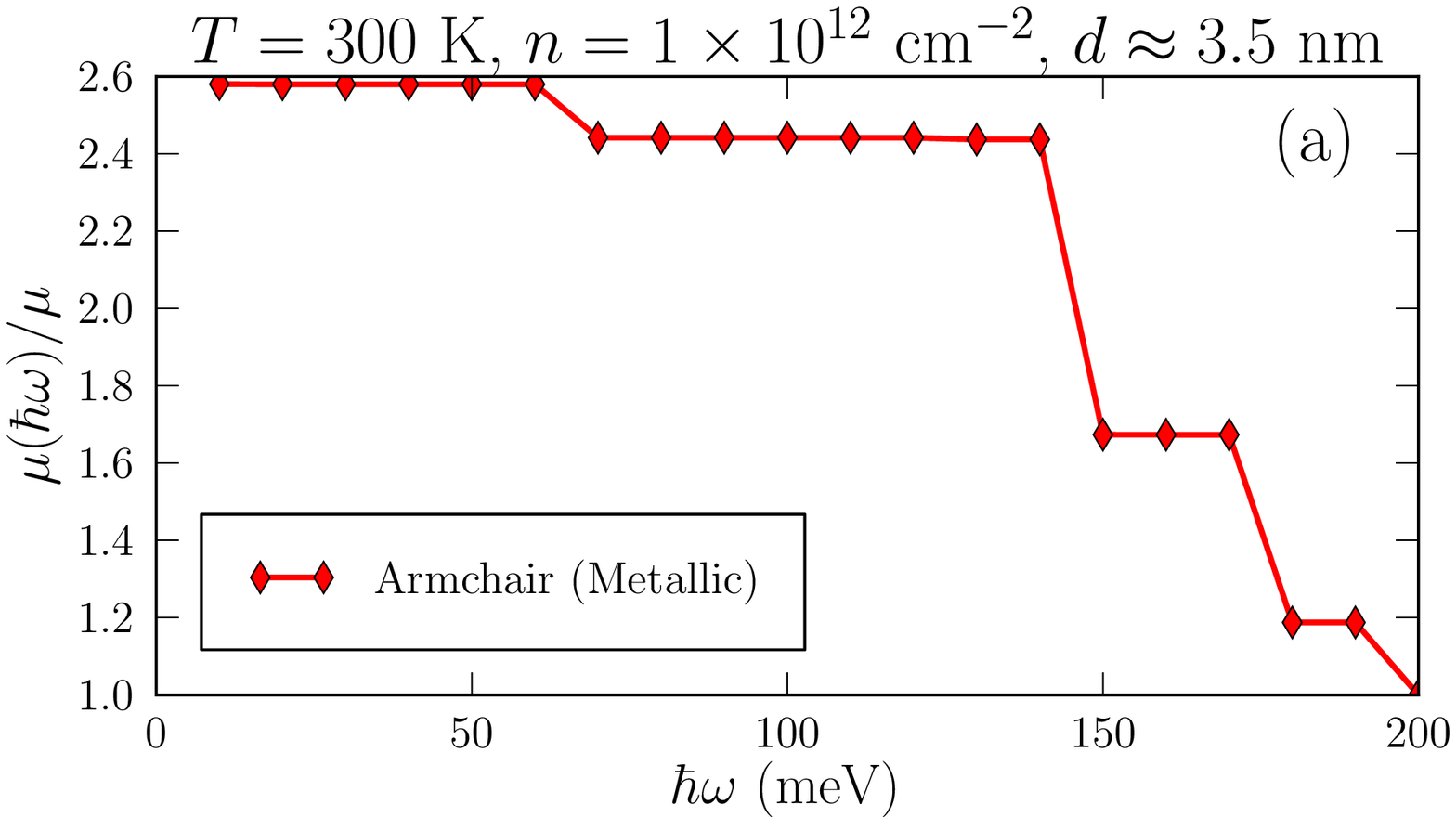}
\includegraphics[width=0.9\columnwidth]{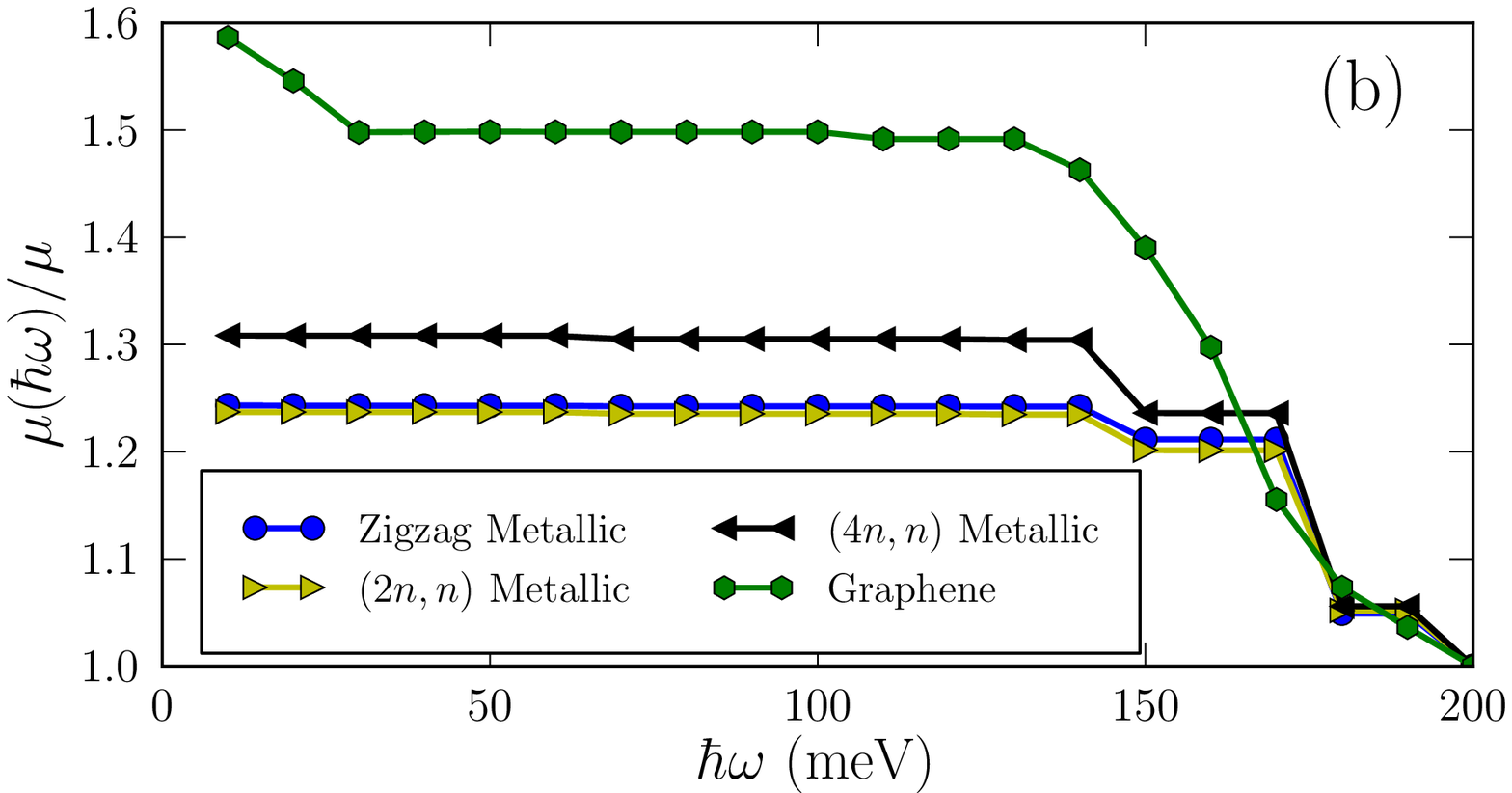}
\includegraphics[width=0.9\columnwidth]{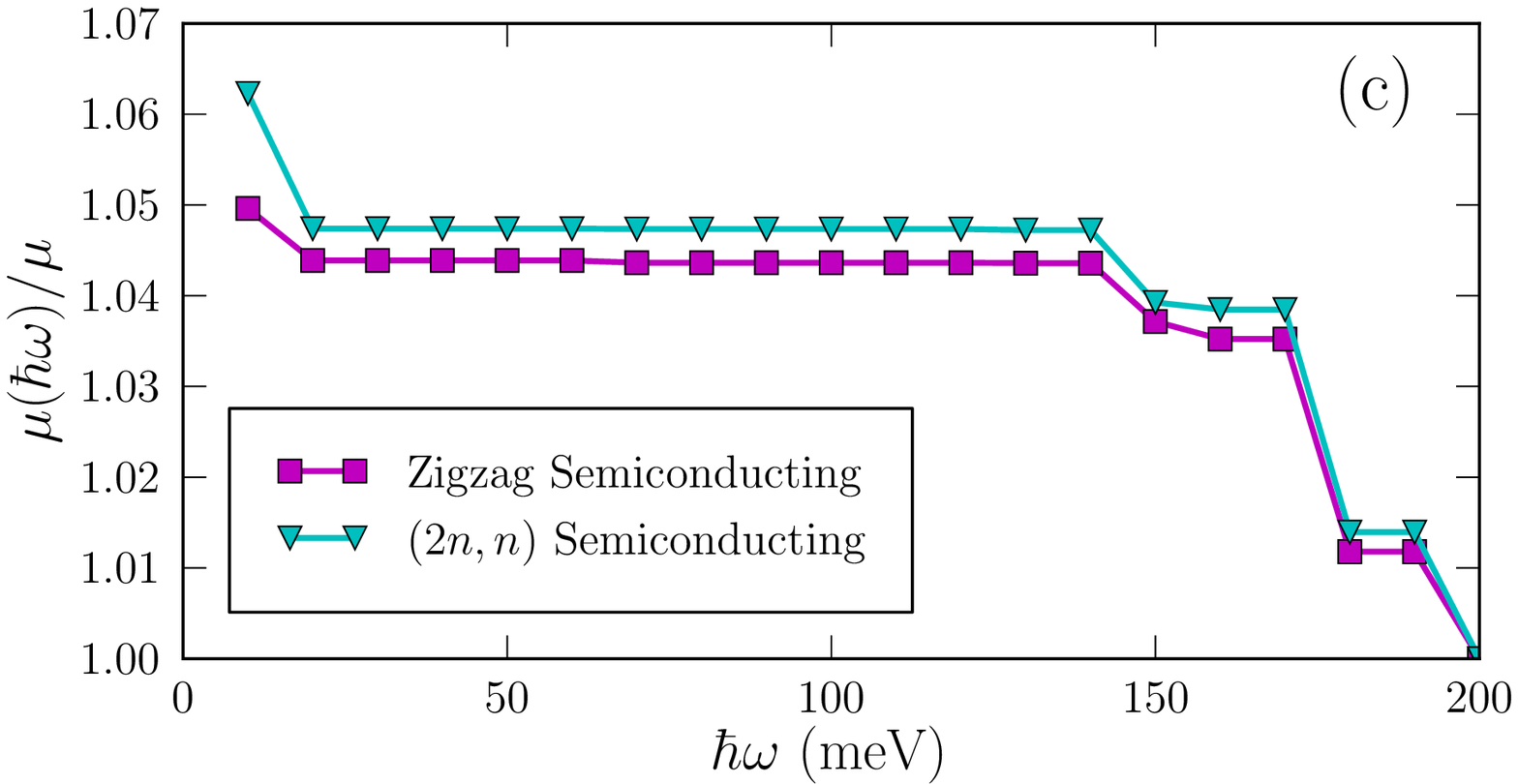}
\caption{Ratio between the mobility calculated by considering phonons with energy up to $\hbar \omega$ and the mobility calculated with all phonons (a) for armchair CNTs; (b) for zigzag, $(2n,n)$, $(4n,n)$ metallic CNTs and graphene; (c) for zigzag, $(2n,n)$ semiconducting CNTs. $d\approx3.5$~nm, $n = 10^{12}$~cm$^{-2}$.}
\label{fig_mu_phlim}
\end{figure}

Figure~\ref{fig_mu_phlim} shows the ratio between the mobility calculated by considering phonons with energy up to $\hbar \omega$ and the mobility including all phonons. In graphene, this ratio has a plateau around 1.5 from $30$ to $130$ meV (green line on Fig.~\ref{fig_mu_phlim}b). This is consistent with previous studies,\cite{Park14} which reported that high-energy phonons contribute to $30\%$ of the resistivity in graphene at room temperature. Similar plateaus are found in CNTs but at different values: 2.5 for armchair CNTs, 1.25 for other metallic CNTs, and 1.05 for semiconducting CNTs ($d\approx3.5$~nm). This means that the relative impact of high energy phonons strongly depends on the type of CNT. In armchair CNTs, this impact is huge -- about $60\%$ of the resistivity -- because the coupling to LA phonons is forbidden in the linear bands and therefore the relative contribution of the optical phonons is more important. The latter is about $20\%$ in other metallic CNTs, and about $5\%$ in semiconducting CNTs, since the coupling to acoustic phonons is typically strong in parabolic bands.

\begin{figure}
\includegraphics[width=0.9\columnwidth]{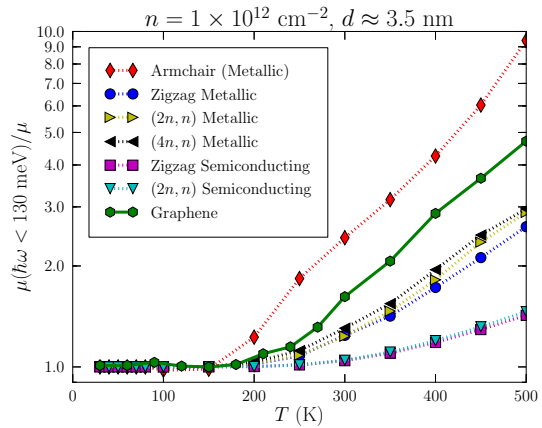}
\caption{Ratio between the mobility calculated by considering phonons with energy below $130$ meV and the mobility calculated with all phonons versus temperature. $d\approx3.5$~nm, $n = 10^{12}$~cm$^{-2}$.}
\label{fig_mu_opt}
\end{figure}

Figure~\ref{fig_mu_opt} shows that 33\% of the electrical resistivity of graphene at 300~K comes from the scattering by high-energy ($>130$~meV) phonons, in agreement with recent {\it ab-initio} calculations.\cite{Park14} This ratio reaches 60\% in 3.5~nm armchair metallic CNTs where the coupling to the LA phonons is weak. On the contrary, this ratio is only 5\% in 3.5~nm semiconducting CNTs because the scattering by low-energy acoustic phonons is very strong and the mobility is low. The coupling to optical phonons also plays an important role in high-field transport in graphene and CNTs.\cite{Yao00,Lazzeri05}
The impact of high-energy phonons increases with the temperature (Fig.~\ref{fig_mu_opt}). The contribution of these phonons to the resistivity reaches 10\% at about $180$, $210$, $250$ and $350$~K in armchair CNTs, graphene, metallic CNTs and semiconducting CNTs, respectively.

\subsubsection{Mobility from CNTs to graphene}

\begin{figure}
\includegraphics[width=0.9\columnwidth]{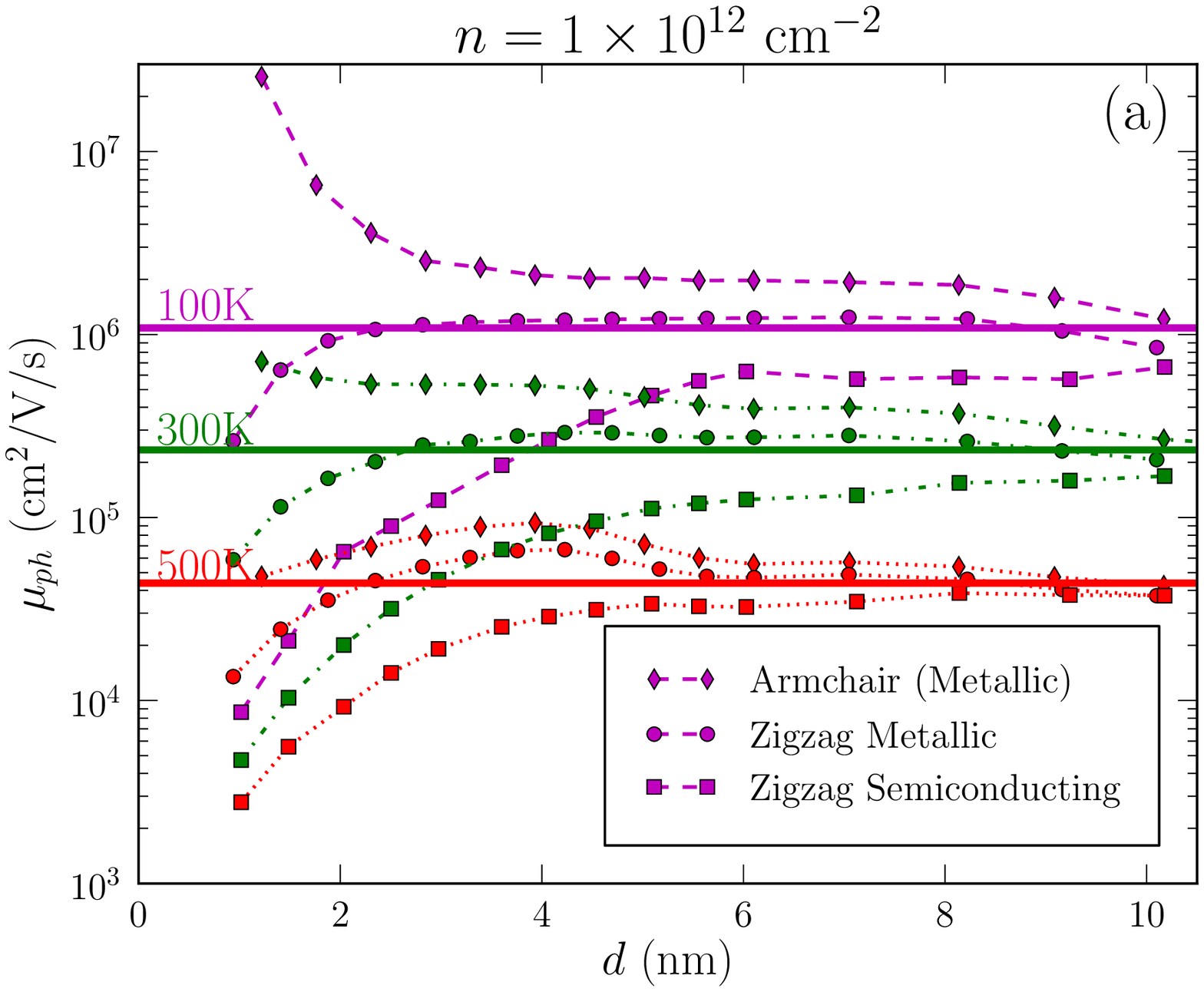}
\includegraphics[width=0.9\columnwidth]{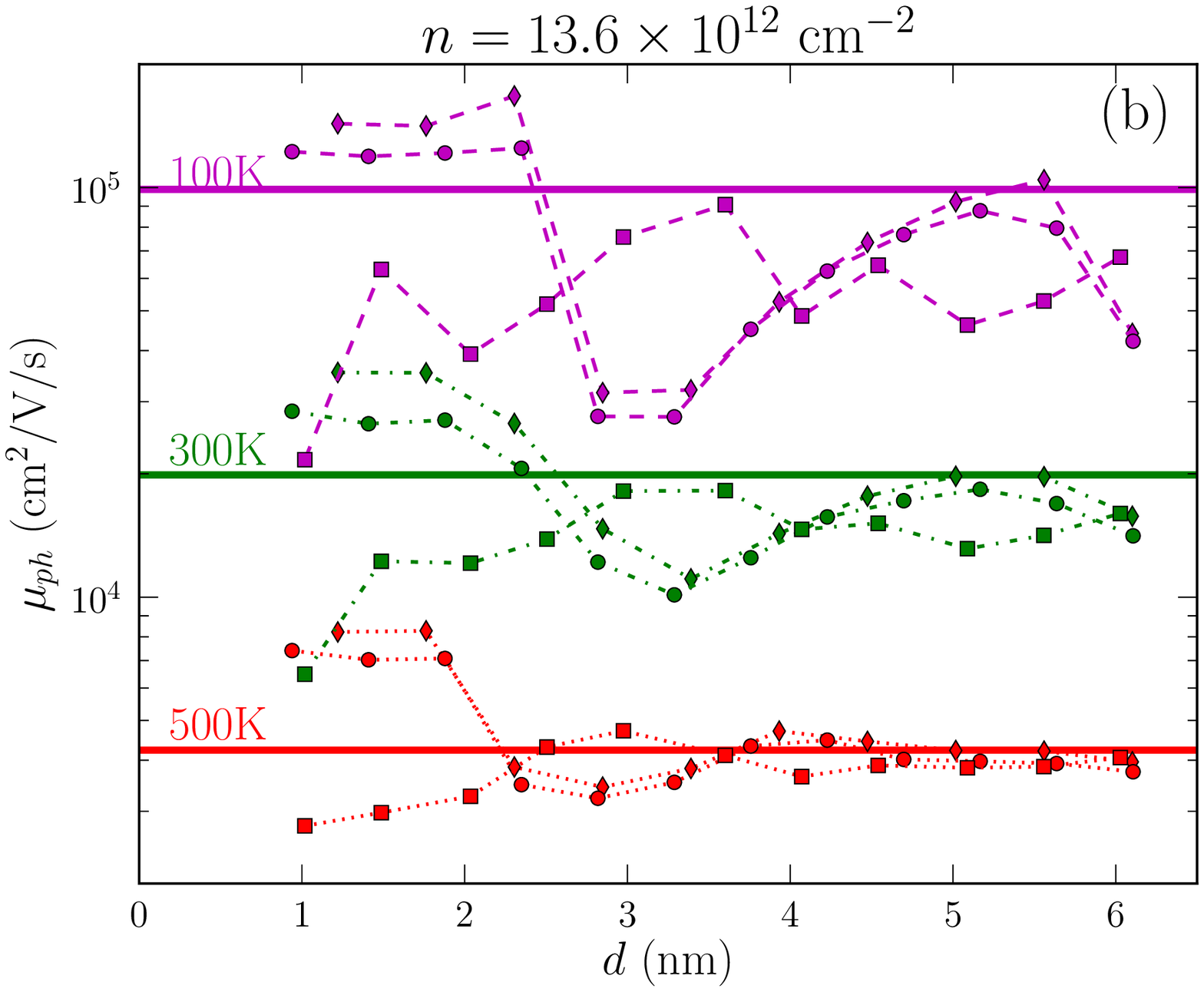}
\caption{Phonon-limited mobility in CNTs versus diameter for $T=100$~K (dashed lines), 300~K (dash-dotted lines) and 500~K (dotted lines). Carrier density: (a) $10^{12}$~cm$^{-2}$, (b) $13.6 \times 10^{12}$~cm$^{-2}$. The horizontal lines indicate the mobility for graphene. The horizontal scale is smaller in (b) than in (a), highlighting the faster convergence of all the mobilities at high carrier density.}
\label{fig_mu_d_T_lowhighn}
\end{figure}

Figure~\ref{fig_mu_d_T_lowhighn} illustrates differently how the mobility in CNTs converges towards the value in graphene when the diameter is increased. The convergence is quite fast at high carrier density ($13.6 \times 10^{12}$~cm$^{-2}$) and high temperature ($T>300$~K, Fig.~\ref{fig_mu_d_T_lowhighn}b). For example, at 500~K, the mobility is basically the same in the three kinds of investigated CNTs as in graphene for $d \approx 5$~nm. The transition from 1D to 2D transport takes place above a certain diameter which strongly depends on temperature and carrier density. Below this threshold, the transport properties strongly depend on the chirality through its effect on the band structure.

\section{Perspectives} 

Experimentally, the diameter of single-walled CNTs which are usually grown is typically between 0.7 and 4~nm. However, a recent work reported the synthesis of single-walled CNTs in the 5--10~nm diameter range (5\% of the CNTs even have a diameter above 10~nm).\cite{Han12} Therefore we believe that the experimental observation of the 1D-2D transition is possible, in particular using the same approach allowing to reach ultrahigh carrier densities up to $4 \times 10^{14}$~cm$^{-2}$ in graphene.\cite{Efetov10}

Another interesting perspective of the present work would be to consider the effects of electron-electron interactions on the 1D-2D transition in CNTs. As discussed above, the transport properties in graphene at high carrier density and temperature ($>20$~K) are unambiguously dominated by electron-phonon scattering. However, theoretical \cite{Egger97,Kane97} and experimental \cite{Bockrath99} studies show that small-diameter metallic CNTs exhibit Luttinger-liquid behaviour characterized by low-energy collective excitations of the electrons. In that case, the tunneling of carriers at energies near the Fermi level is strongly suppressed, so that the conductance increases as a power law with respect to temperature.\cite{Bockrath99} On the contrary, the conductance in semiconducting CNTs strongly decreases with temperature, as expected for diffusive transport limited by phonons.\cite{Zhou05} All these works deal with small-diameter CNTs at low carrier density. Therefore it would be extremely interesting to extend experimental studies on CNTs to larger diameter and higher carrier density. The electron-phonon interactions should indeed overcome electron-electron interactions when the number of populated subbands increases, in particular for parabolic bands. In these conditions, the carrier mobility in metallic CNTs should follow our predictions above certain thresholds that must be found.

\section{Conclusion}

It is shown that the phonon-limited mobility in 1D CNTs approaches that of 2D graphene continuously by increasing the size of CNTs, the carrier density, or the temperature. The physics of this transition has been studied using atomistic calculations combining a tight-binding model for electrons, a force-constant model for phonons, the computation of all the electron-phonon couplings, and a full resolution of the Boltzmann transport equation. This approach gives carrier mobility in graphene in excellent agreement with experiments \cite{Efetov10} and DFT calculations,\cite{Park14} and therefore can be used to predict the mobility in CNTs in a wide range of configurations. The mobility in CNTs can be higher or smaller than in graphene depending on chirality, diameter, carrier density or temperature but converges to the same value above varying thresholds.
This 1D to 2D transition takes place when the number of subbands situated in the transport energy window is sufficiently large to smooth out the effect of the 1D confinement on the band structure and on the electron-phonon coupling.

%
%
\appendix

\section{Parameters for the 4NN force constant model}
\label{4NN_model}

We have refitted the force constant method of Ref.~\onlinecite{Wirtz04} to 
an \textit{ab-initio} calculation of the phonon-dispersion of graphene
using the \texttt{ABINIT} code.\cite{Gonze09} 
The electronic structure is calculated in the local-density approximation (LDA) using a regular 60x60 reciprocal mesh in the first Brillouin zone and an energy cut-off of 35 Ha using a LDA functional. A thermal Fermi-Dirac smearing of 0.002 Ha is employed. We find the optimized cell parameter to be 4.631 \AA. The dynamical matrices were calculated using density functional perturbation theory (DFPT) on a 30x30 q-mesh. Since LDA overbinds, i.e., phonon frequencies have the tendency to be slightly too high, a scaling factor is used such that the phonon frequencies of the LO/TO mode at Gamma match the experimental value.\cite{Wirtz04}

The (real-space) force constants between two particular atoms $a$ and $b$ are defined as the second derivatives of the total energy of the system with respect to the displacements of atom $a$ in direction $i$ and of atom $b$ in direction $j$:

\begin{equation}
C_{ij}^\prime= \frac{\partial^2 E}{\partial x^a_i\partial x^b_j}. 
\end{equation}

In local coordinates, direction $x_1$ is along the line connecting the two atoms, $x_2$ is perpendicular to this line in the plane of the graphene sheet, $x_3$ is the out-of-plane direction. In this local reference frame we 
define the longitudinal forces, ($\phi_n^l$), transverse in-plane ($\phi_n^{ti}$), and  transverse-out-of-plane ($\phi_n^{to}$) forces that act on a particular atom when its $n$th nearest-neighbor is displaced. In the conventional 4NN-force constant model, only these ``diagonal'' terms are fitted. In our model, we include also the ``off-diagonal'' coupling between the longitudinal direction
and the transverse-in-plane direction ($\epsilon_n^{l/ti}$ and $\epsilon_n^{ti/l}$). The force-constant matrix for the interaction between two atoms in the 
local reference frame thus reads:

\begin{equation}
\mathbf{C}_n^\prime = 
\begin{pmatrix}
  \phi_n^l        & \epsilon_n^{l/ti} &           0 \\
\epsilon^{ti/l}_n &       \phi_n^{ti} &           0 \\
                0 &                 0 & \phi_n^{to} \\
\end{pmatrix}
\end{equation}

The off-diagonal force constants $\epsilon_n^{l/ti}$ and $\epsilon_n^{ti/l}$ obey the following relations:\cite{Dubay}

\begin{eqnarray}
\epsilon_1^{l/ti} = \epsilon_3^{l/ti} = 0 && ~~~~~~ \epsilon_2^{l/ti} = -\epsilon_2^{ti/l} \nonumber \\
\epsilon_1^{ti/l} = \epsilon_3^{ti/l} = 0 && ~~~~~~ \epsilon_4^{l/ti} = \epsilon_4^{ti/l}
\end{eqnarray}

\begin{figure}
\includegraphics[width=0.98\columnwidth]{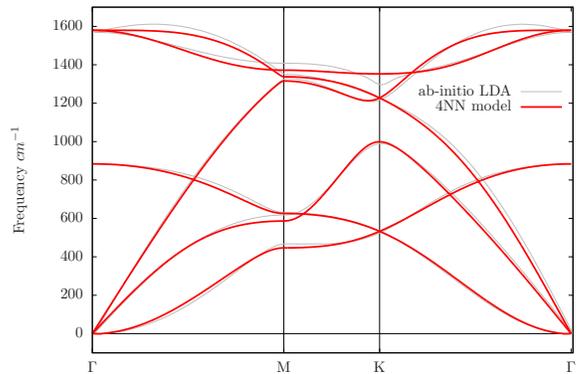}
\caption{Phonon dispersion of graphene. The red lines are the frequencies calculated using the 4NN force constant model, and grey lines are DFT-LDA calculations using \texttt{ABINIT}.}
\label{gra_dispersion}
\end{figure}

During the fitting, we have noticed that the off-diagonal terms of the second and the fourth nearest neighbor ($\epsilon_2^{l/ti}$ and $\epsilon_4^{l/ti}$) are essential. While a fairly decent fit of the phonon-dispersion alone can be achieved without these off-diagonal terms, the correct ratio of the amplitudes of optical and acoustic phonon components in the transverse and longitudinal acoustic branches at $q \ne 0$ can only be achieved with the inclusion of the off-diagonal forces. By using the parameters in Table ~\ref{4NN_para}, the \textit{ab-initio} ratio  was reproduced to a very good degree. The comparison of the phonon dispersion relations between the \textit{ab-initio} calculation and the 4NN force constant model is shown in Fig. ~\ref{gra_dispersion}. We note that the agreement is good but the 4NN model cannot reproduce the Kohn anomalies in the two highest-optical branches. This would require the inclusion of many more distant neighbor interactions in the model (even infinitely many, if one wants to reproduce the kink\cite{Dubay}). However, this does not seem to be necessary for our present study since the 4NN model gives electron-phonon scattering rates and carrier mobilities in excellent agreement with \textit{ab-initio} calculations for graphene.\cite{Park14}


\begin{table}[!htb]
\begin{center}
\begin{tabular}{ c c c c c }
\hline
\hline
n & 1 & 2 & 3 & 4\\
\hline
$\phi_n^l$ ($10^4$ dyn/cm) &  40.905 & 7.402  & -1.643  & -0.609\\
                                                   
$\phi_n^{ti}$ ($10^4$ dyn/cm) &  16.685 & -4.051  & 3.267  &  0.424\\
                                                  
$\phi_n^{to}$ ($10^4$ dyn/cm) &  9.616 & -0.841 &  0.603 & -0.501\\

$\epsilon_n^{l/ti}$ ($10^4$ dyn/cm) &  0.000 &   0.632 & 0.000 & -1.092\\

$\epsilon_n^{ti/l}$ ($10^4$ dyn/cm) &  0.000 &  -0.632 & 0.000 & -1.092\\
\hline
\end{tabular}
\end{center}
\caption{Parameters of the 4NN force constant model with the off-diagonal couplings. The corresponding dispersion relation is shown in Fig.~\ref{gra_dispersion}.}
\label{4NN_para}
\end{table}

\section{Resistivity of graphene including surface optical phonon scattering}
\label{resistivity_graphene}

\begin{figure}
\includegraphics[width=0.90\columnwidth]{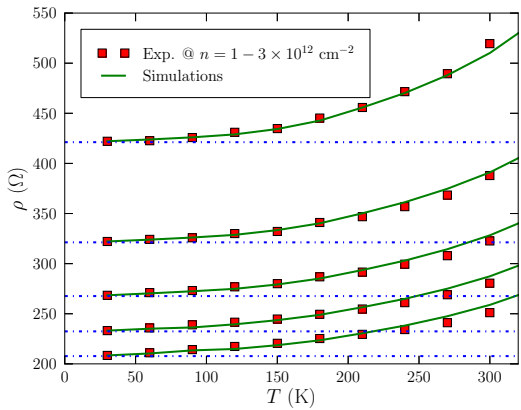}
\caption{Comparison between our simulations (green dotted lines) and experimental data\cite{Zou10} (red markers) for the resistivity of graphene at carrier density varying from $1$ (top) to $3\times 10^{12}$~cm$^{-2}$ (bottom) with a step of $0.5\times10^{12}$~cm$^{-2}$. The blue horizontal dotted lines represent the impurity-limited resistivity $\rho_0$.}
\label{gra_exp}
\end{figure}

The resistivity of graphene was measured in Ref.~\onlinecite{Zou10} for carrier densities varying from $1$ to $3\times 10^{12}$~cm$^{-2}$. In the following, we show that this resistivity consists of three major components:

\begin{equation}
\label{eq:resistance}
\rho(T) = \rho_0+\rho_{ph}(T)+\rho_{SO}(T),
\end{equation}

\noindent where $T$ is temperature, $\rho_0$ is due to impurities, $\rho_{ph}$ is due to the intrinsic phonons of graphene, and $\rho_{SO}$ is due to the surface optical phonons of the substrate. $\rho_{ph}$ is calculated as described in the main document. $\rho_{SO}$ is calculated with the model of Ref.~\onlinecite{Konar10} in which SiO$_{2}$ is considered for the substrate. $\rho_0$ is determined by taking the difference between experimental data and the sum of $\rho_{ph}$ and $\rho_{SO}$ at $30$~K. Figure~\ref{gra_exp} compares the results of the calculation with experimental data \cite{Zou10} in a wide range of temperature and for different electron densities $n_{2D}$. It is clear that they agree very well with each other.

\begin{figure}
\includegraphics[width=0.90\columnwidth]{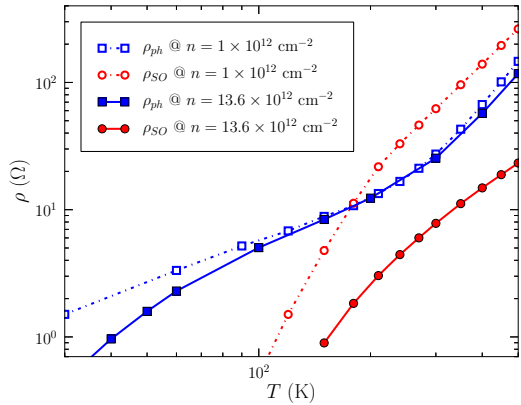}
\caption{$\rho_{ph}$ (blue makers) and $\rho_{SO}$ (red markers) versus temperature at electron densities $1$ (empty markers) and $13.6\times 10^{12}$~cm$^{-2}$ (filled markers).}
\label{t_dept}
\end{figure}

Figure~\ref{t_dept} shows the temperature dependence of $\rho_{ph}$ and $\rho_{SO}$. At low temperature, the effect of the surface optical phonon scattering is not significant because the lowest surface optical phonon energy considered in the model is $59.98$~meV. The surface optical phonon scattering is important at high temperature when the carrier density is low. It almost dominates the resistivity at $n_{2D} = 10^{12}$~cm$^{-2}$. However, at higher densities, its contribution becomes negligible due to strong screening.\cite{Konar10}

\section{Band structures of CNTs}
\label{band_CNT}

\begin{figure*}

\includegraphics[width=0.49\columnwidth]{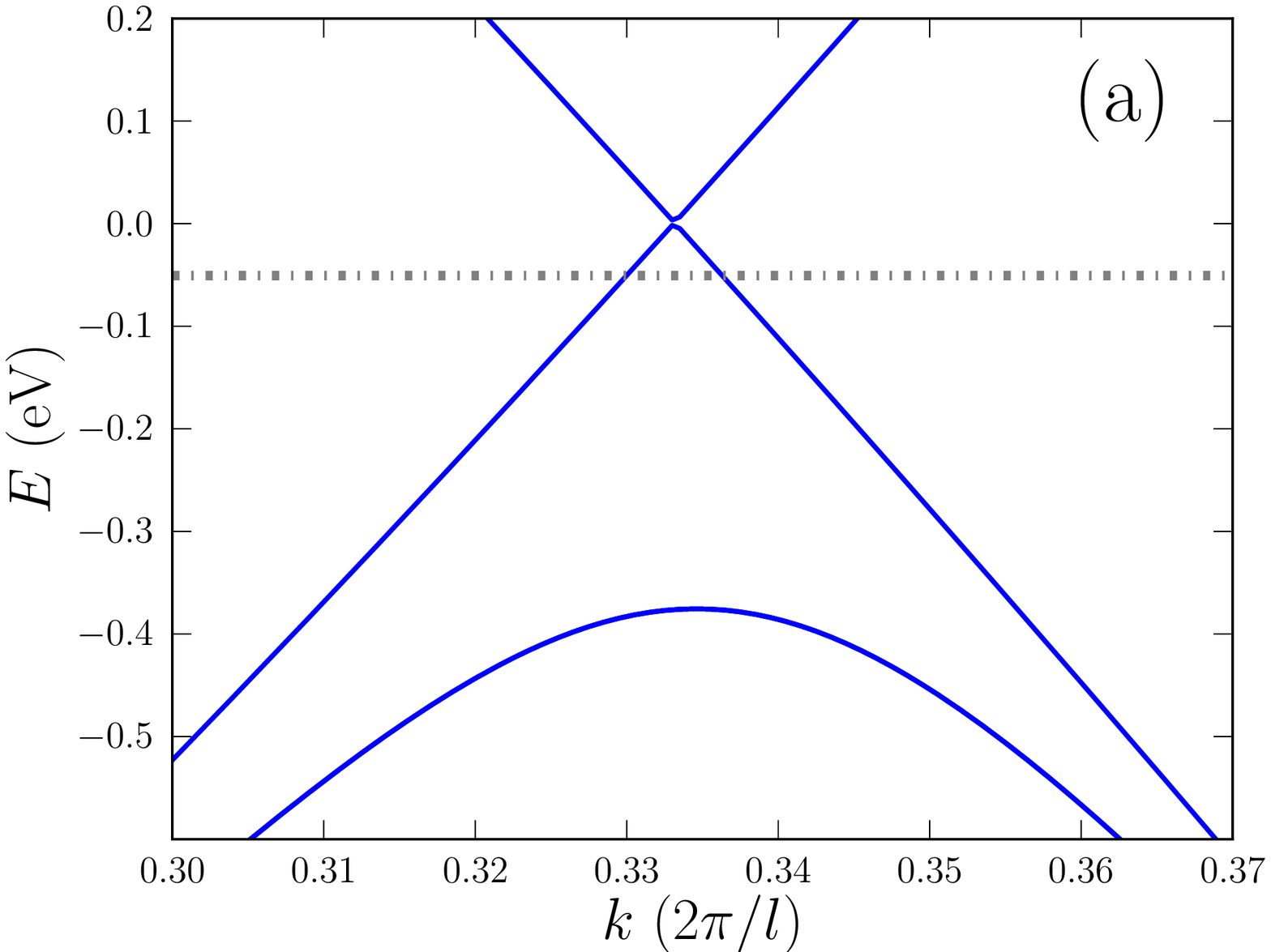}
\includegraphics[width=0.49\columnwidth]{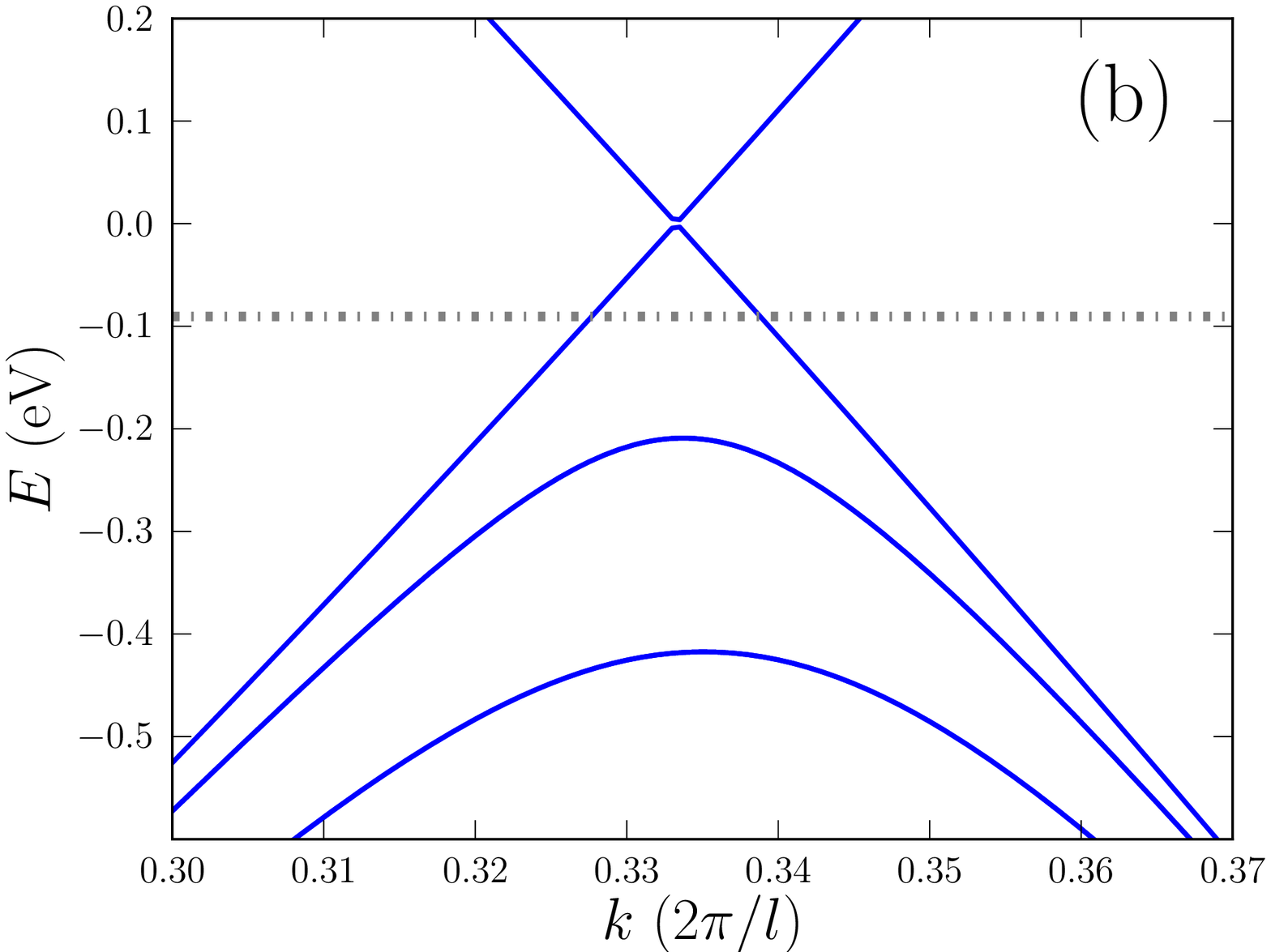}
\includegraphics[width=0.49\columnwidth]{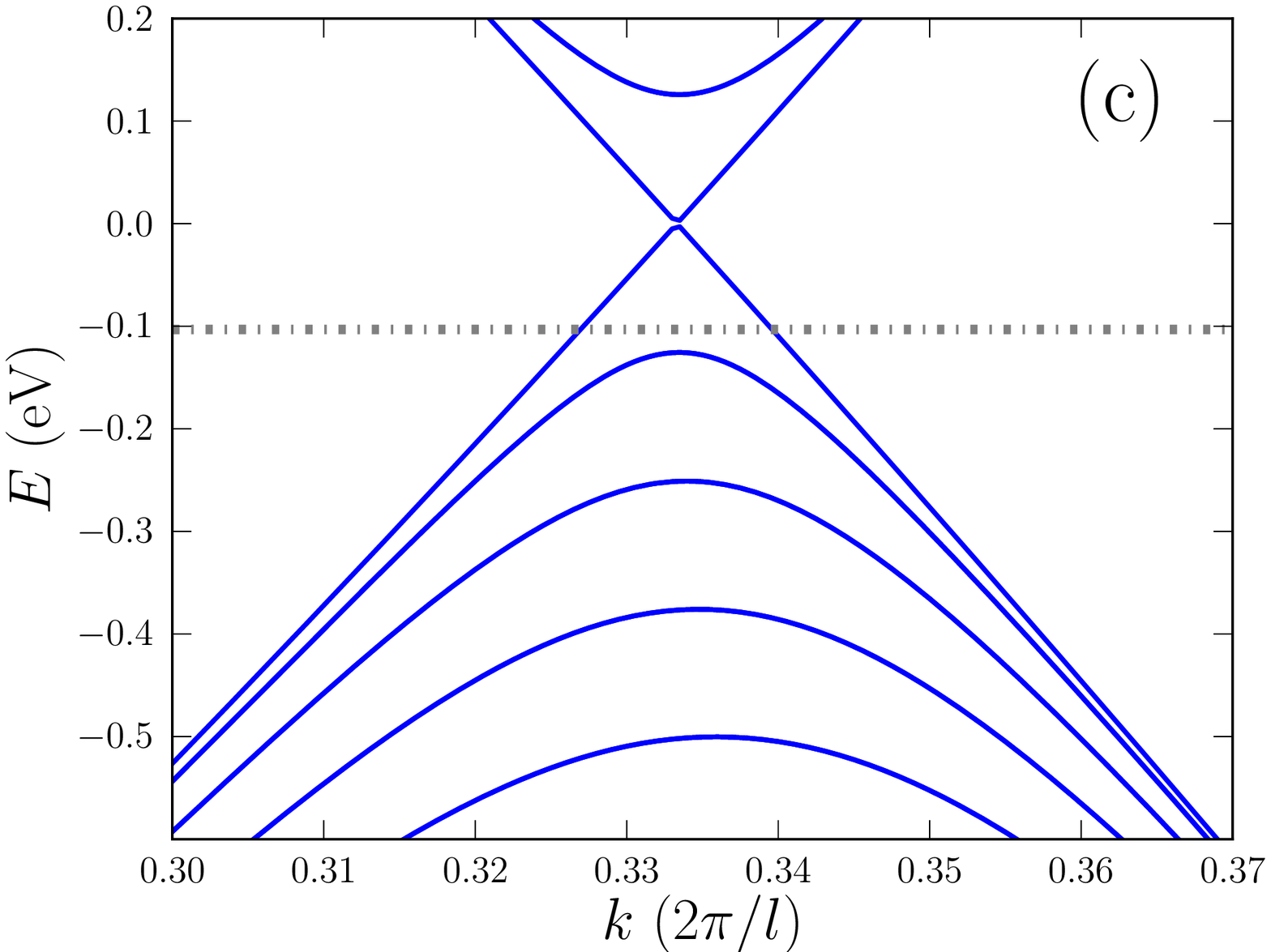}
\includegraphics[width=0.49\columnwidth]{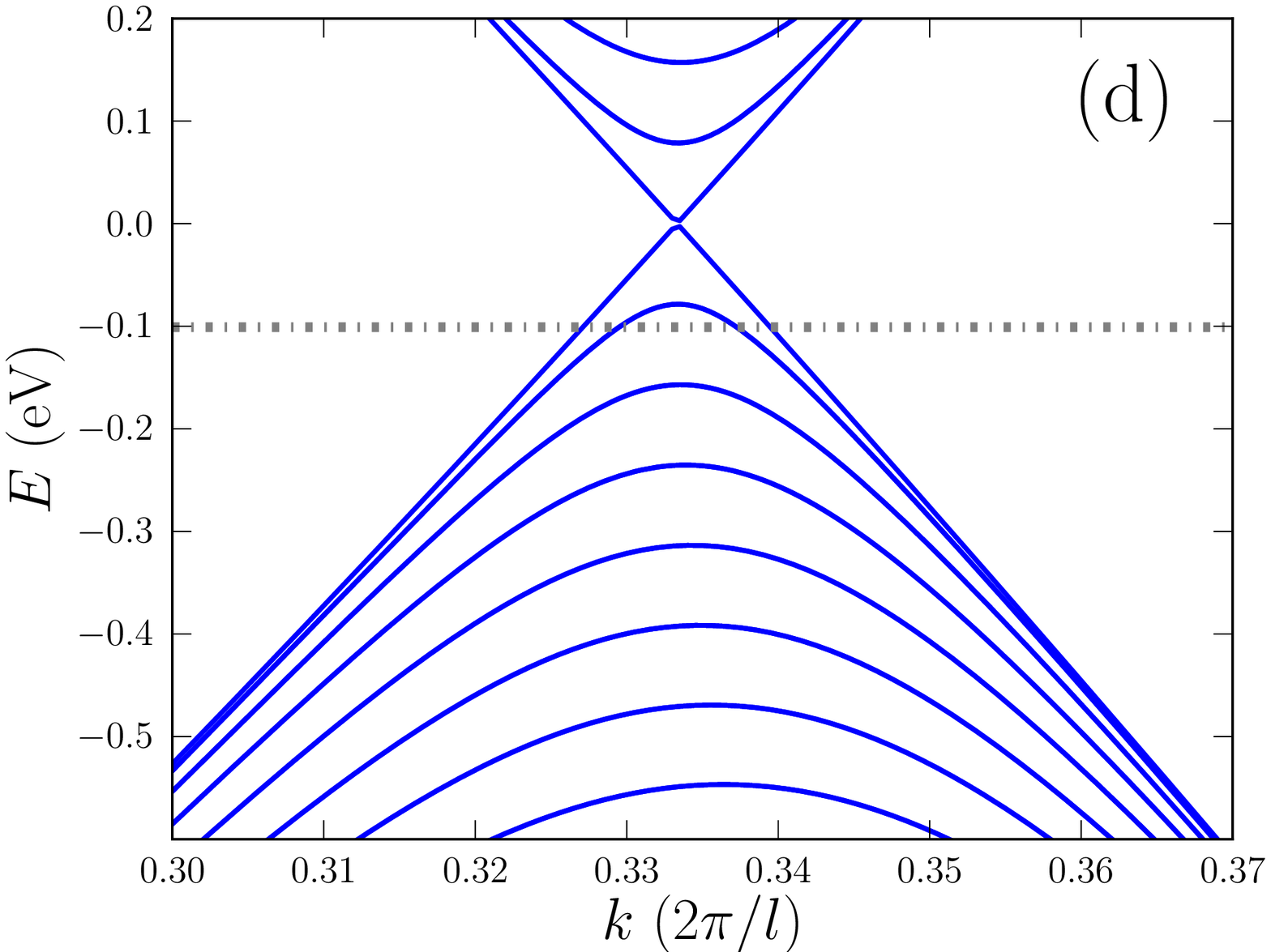}\\
\includegraphics[width=0.49\columnwidth]{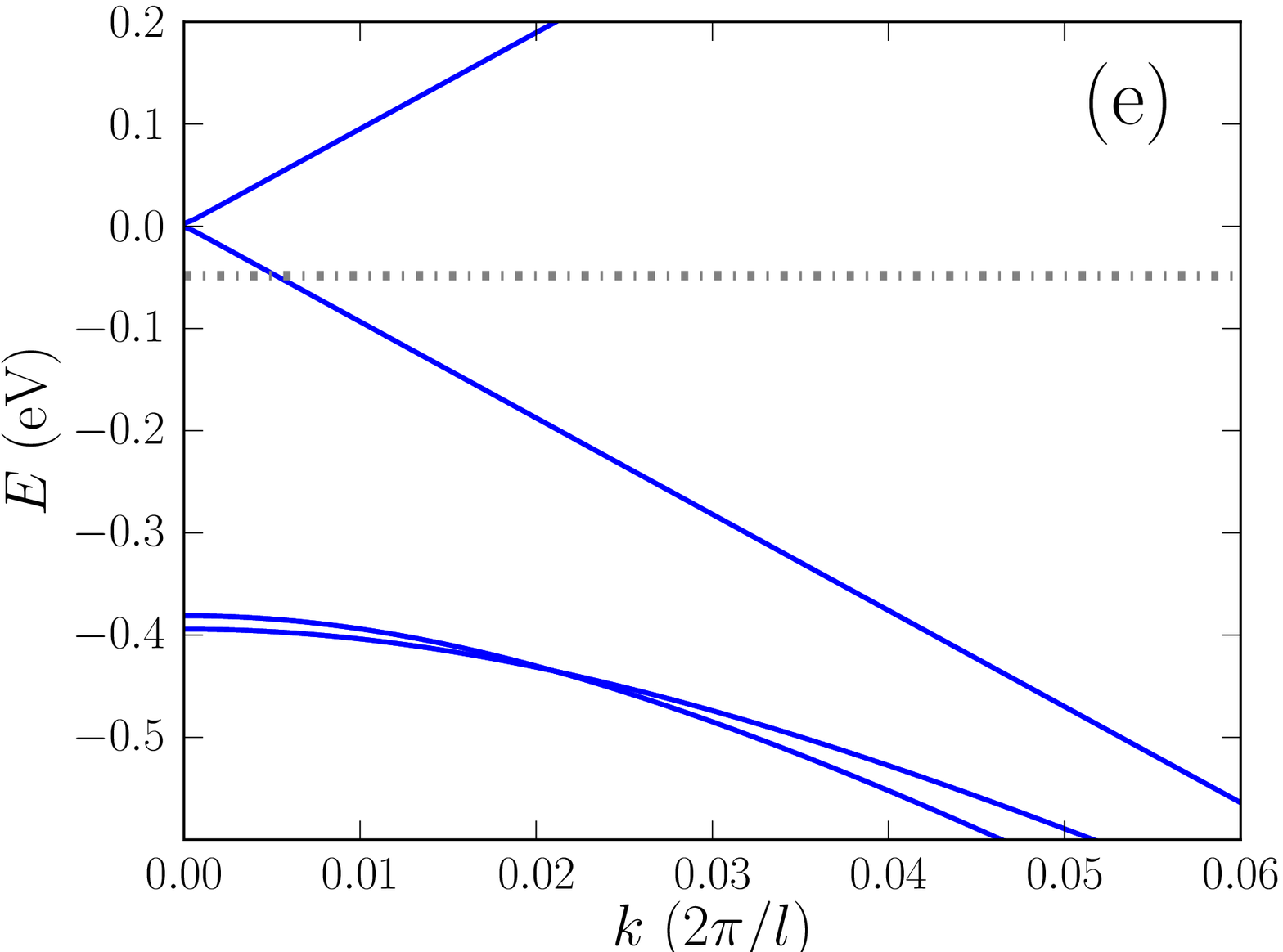}
\includegraphics[width=0.49\columnwidth]{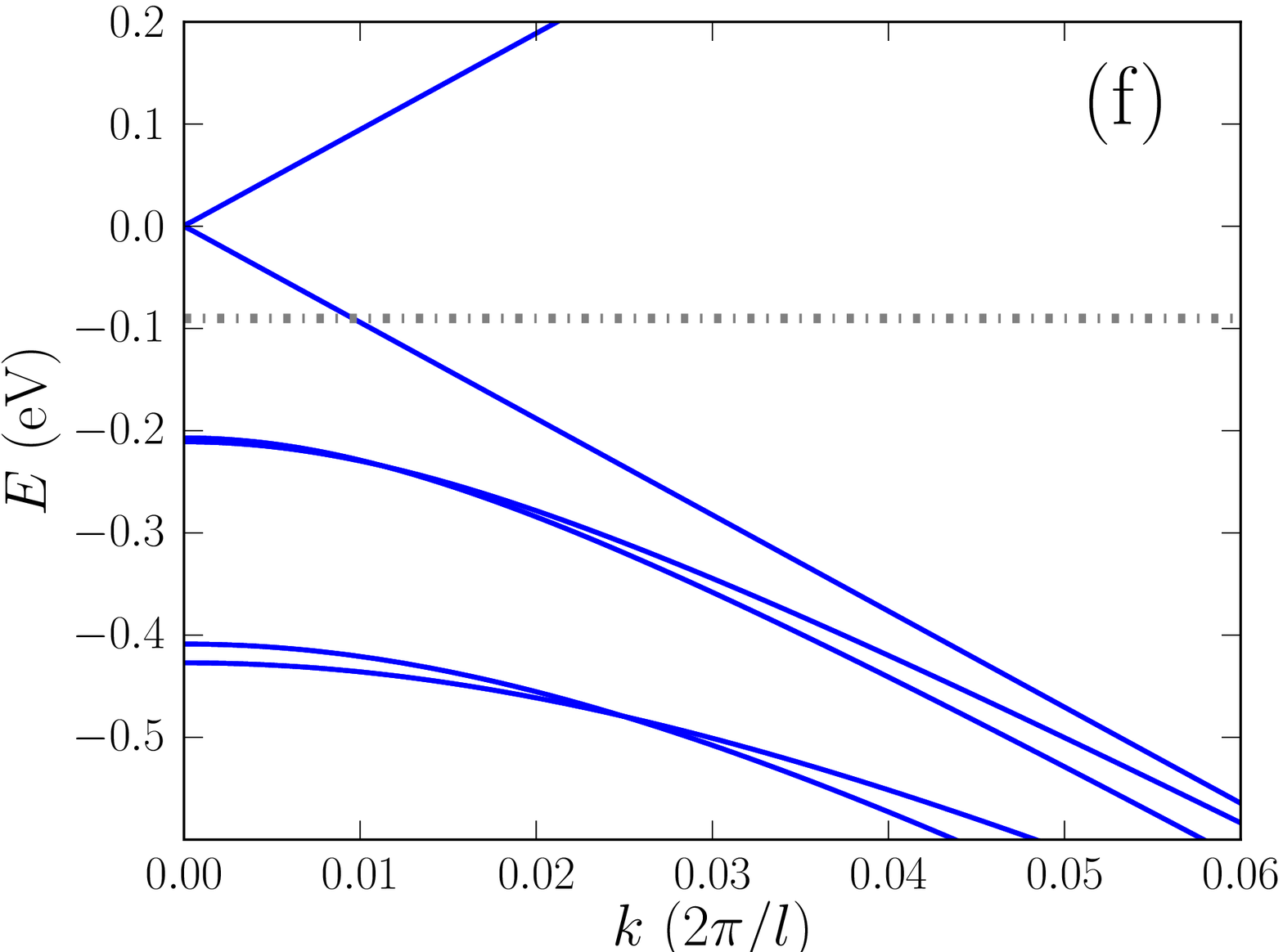}
\includegraphics[width=0.49\columnwidth]{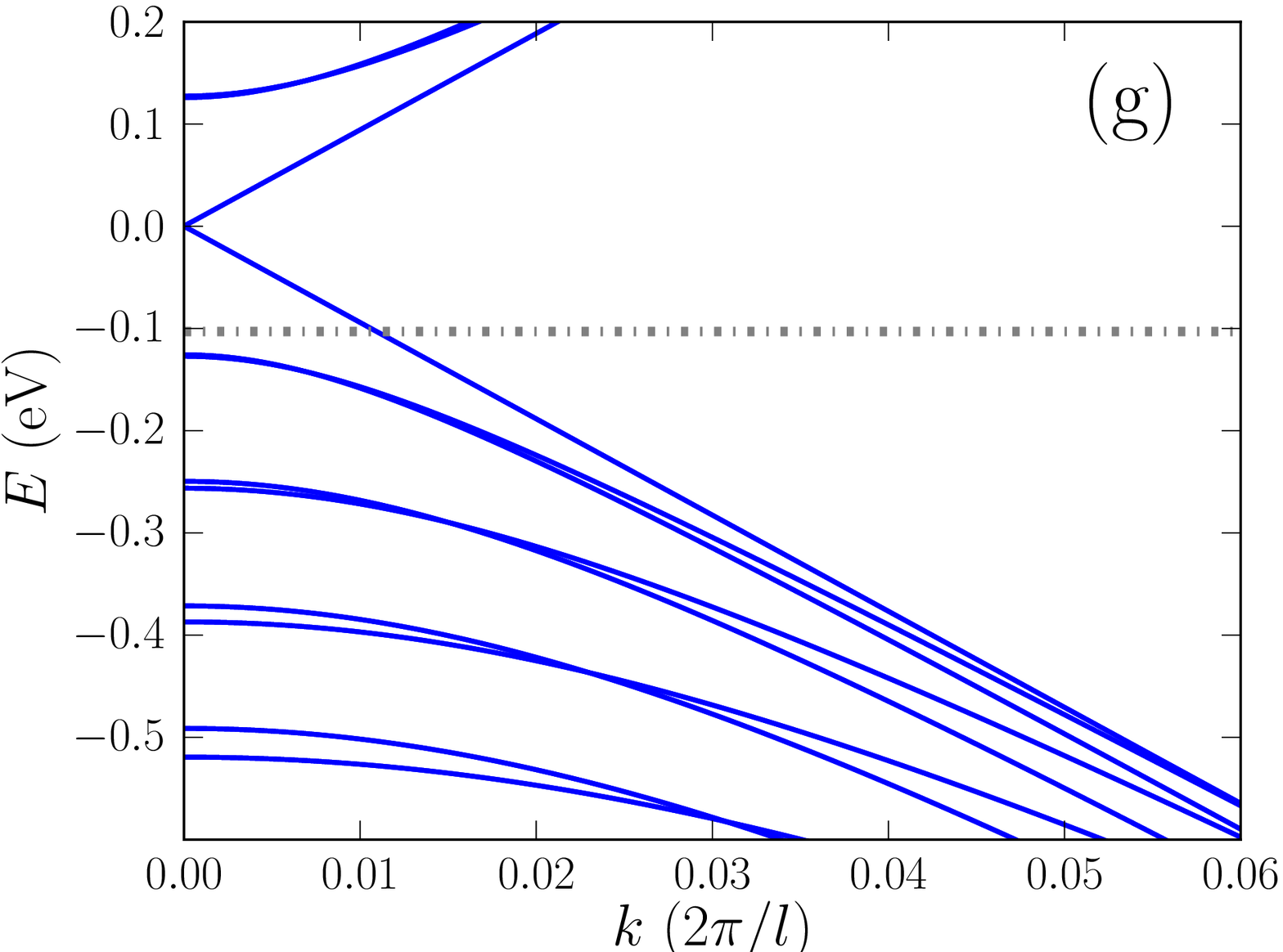}
\includegraphics[width=0.49\columnwidth]{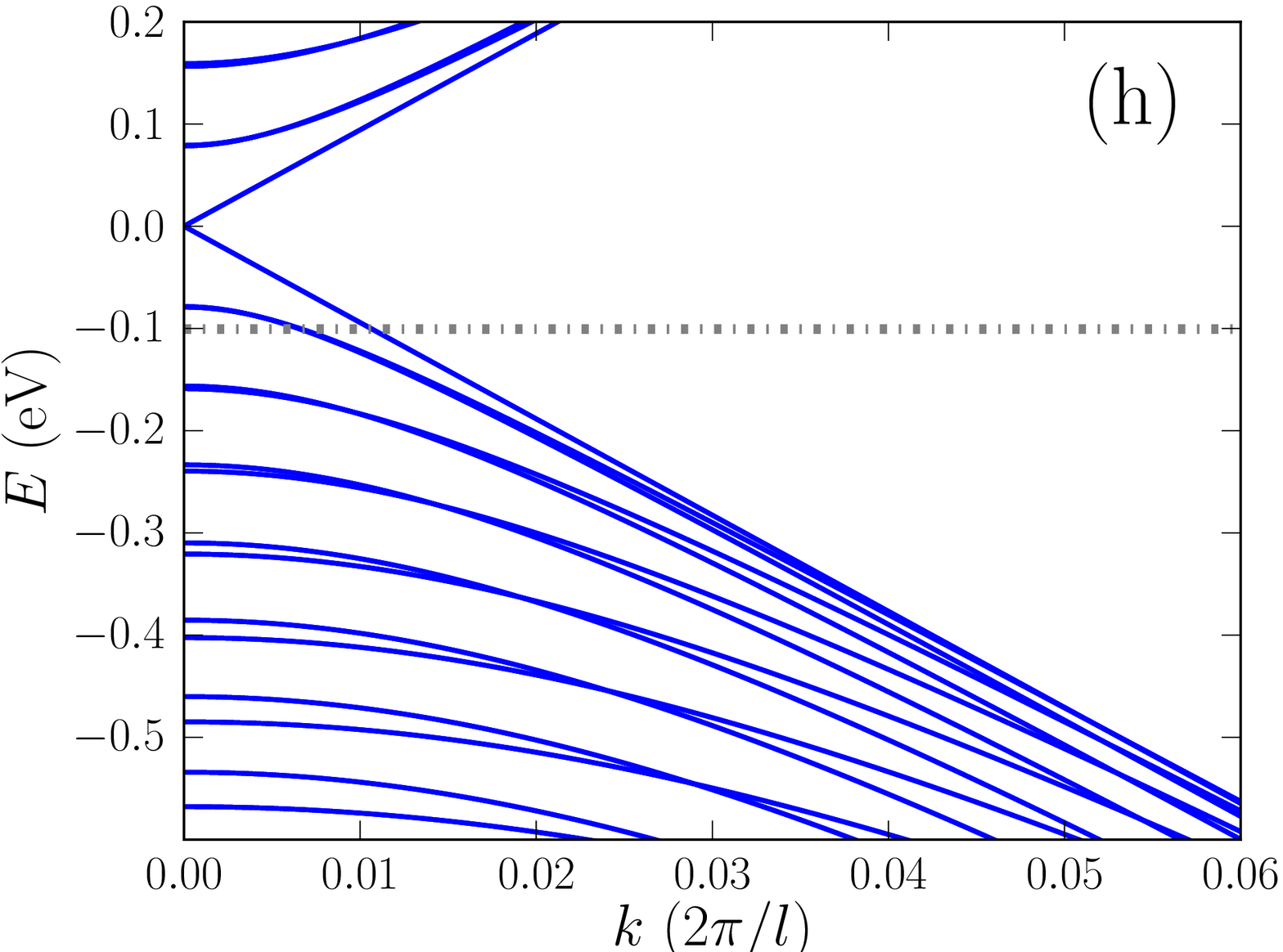}\\
\includegraphics[width=0.49\columnwidth]{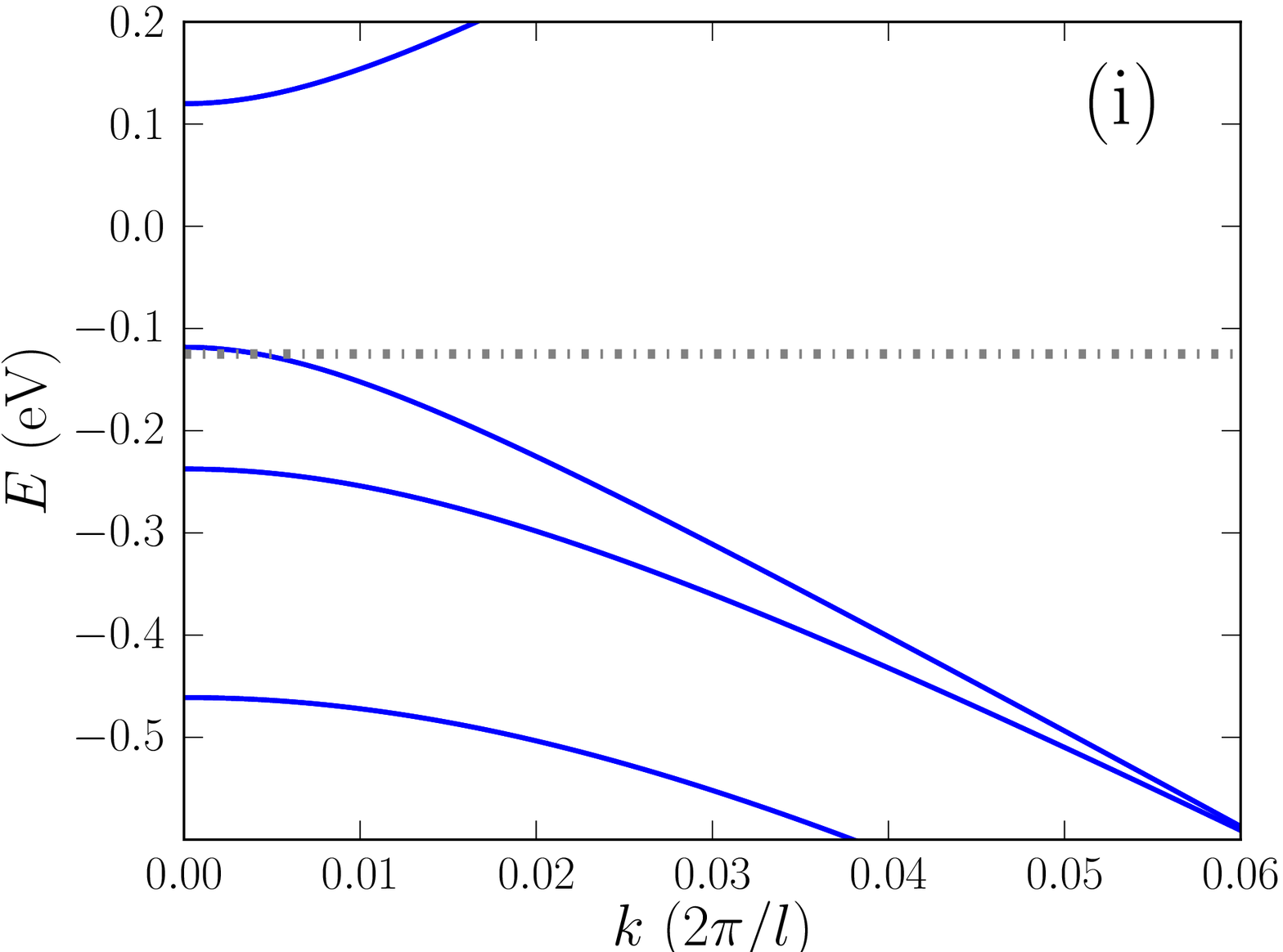}
\includegraphics[width=0.49\columnwidth]{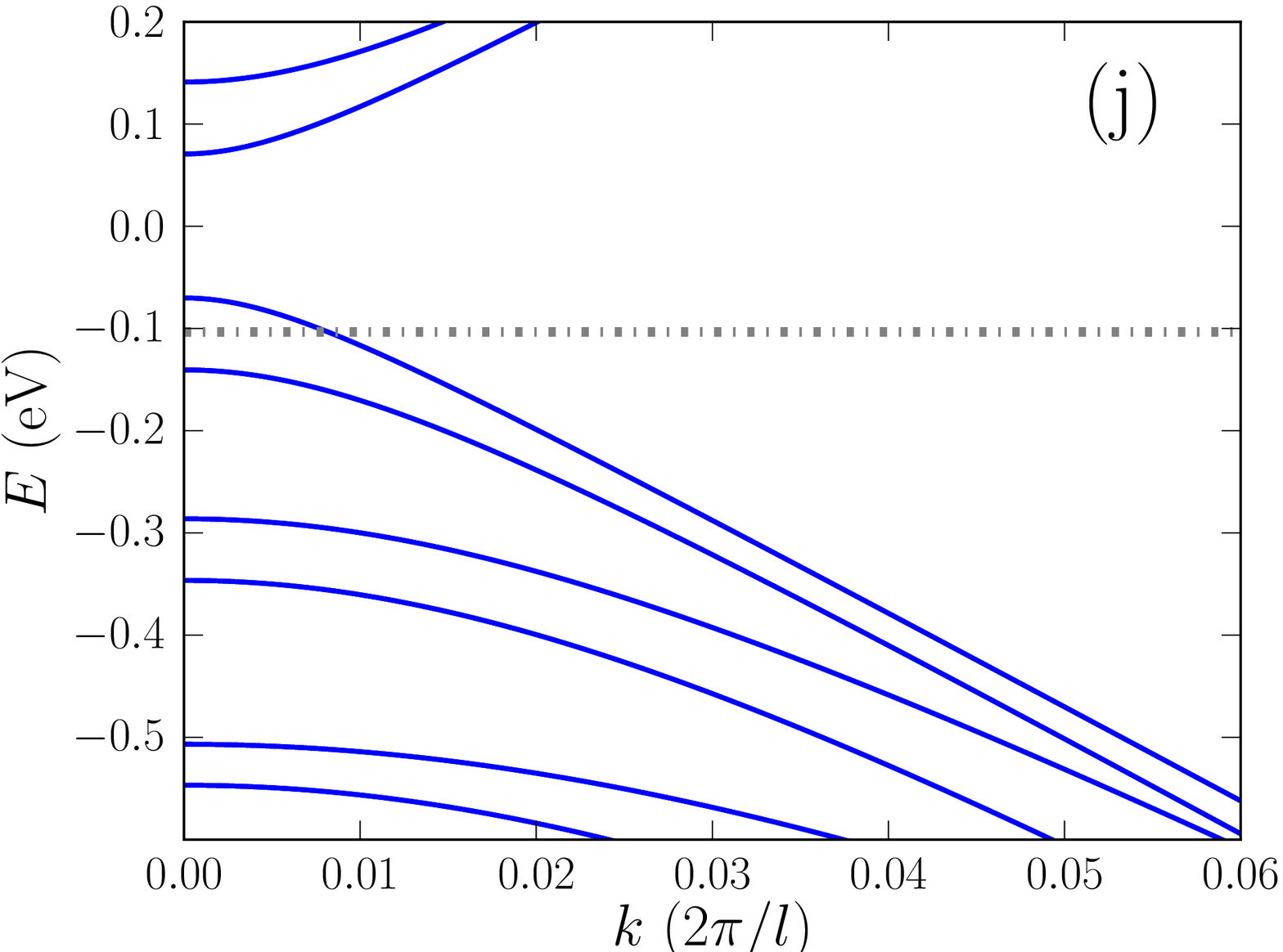}
\includegraphics[width=0.49\columnwidth]{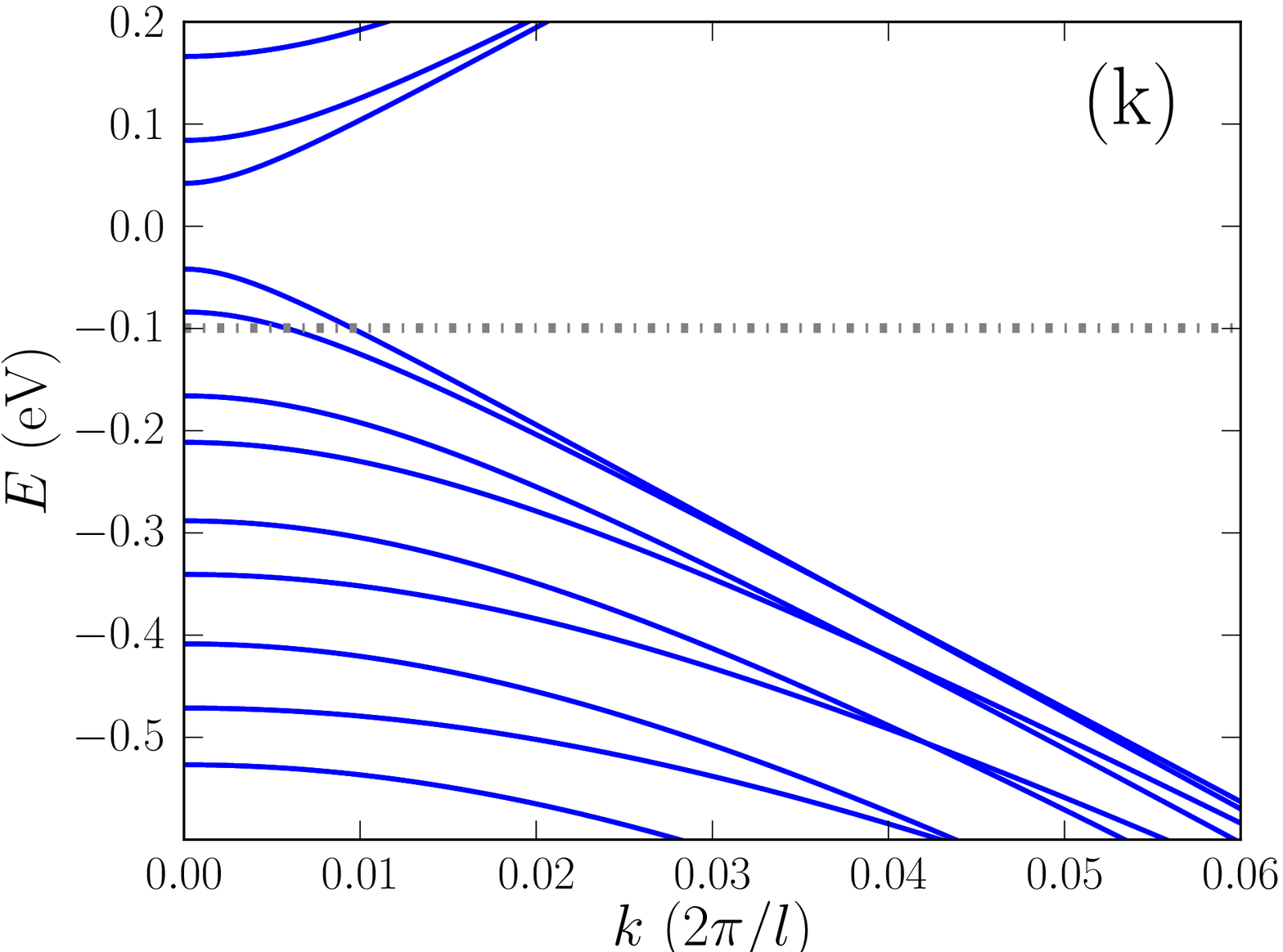}
\includegraphics[width=0.49\columnwidth]{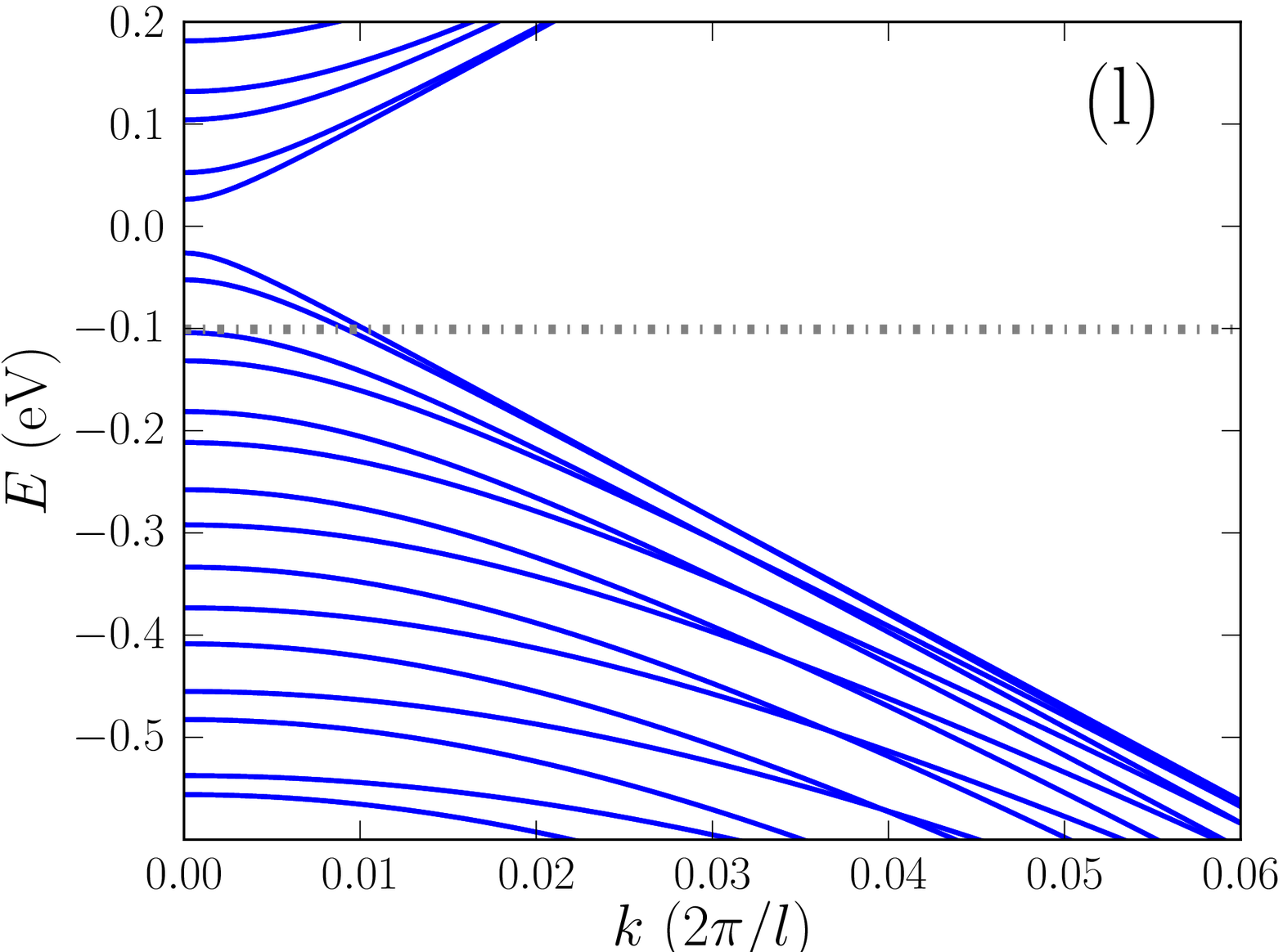}
\caption{Band structures of metallic armchair (a-d), metallic zigzag (e-h) and semiconducting zigzag (i-l) CNTs with diameter of $3.5$ (a,e,i), $6$ (b,f,j), $10$ (c,g,k) and $16$~nm (d,h,l). The 2D effective carrier density is fixed at $10^{12}$~cm$^{-2}$. The horizontal line represents the Fermi level at $300$~K.}
\label{fig_mu_d_300K_low_n_size}
\end{figure*}

Figure~\ref{fig_mu_d_300K_low_n_size} shows the band structure of metallic armchair, metallic zigzag and semiconducting zigzag CNTs for selected diameters. For small diameter CNTs, only the lowest band is involved in the transport. The effective mass of semiconducting CNTs decreases with increasing diameter, leading to higher mobility in CNTs with large diameter.

\begin{figure*}

\includegraphics[width=0.49\columnwidth]{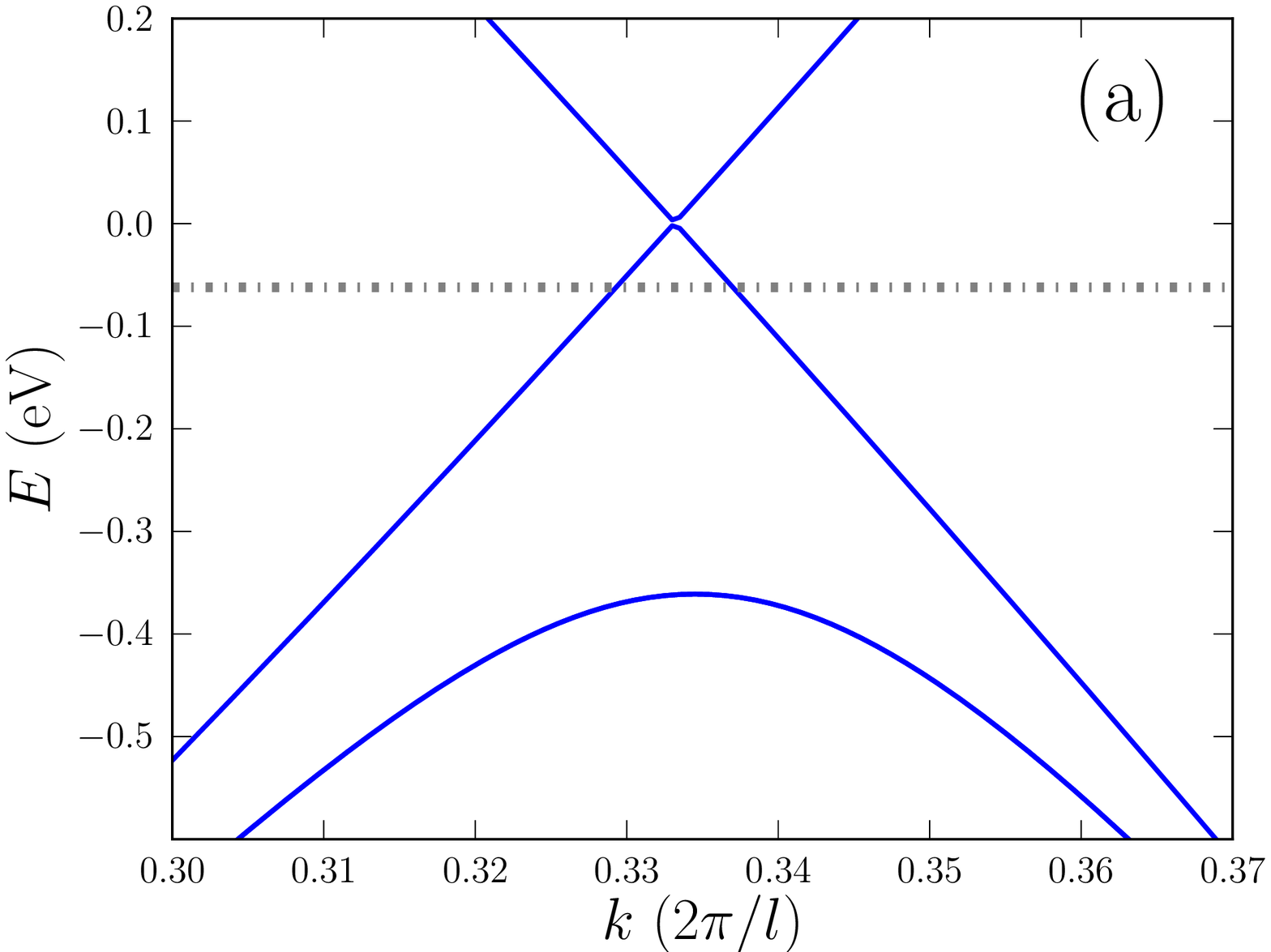}
\includegraphics[width=0.49\columnwidth]{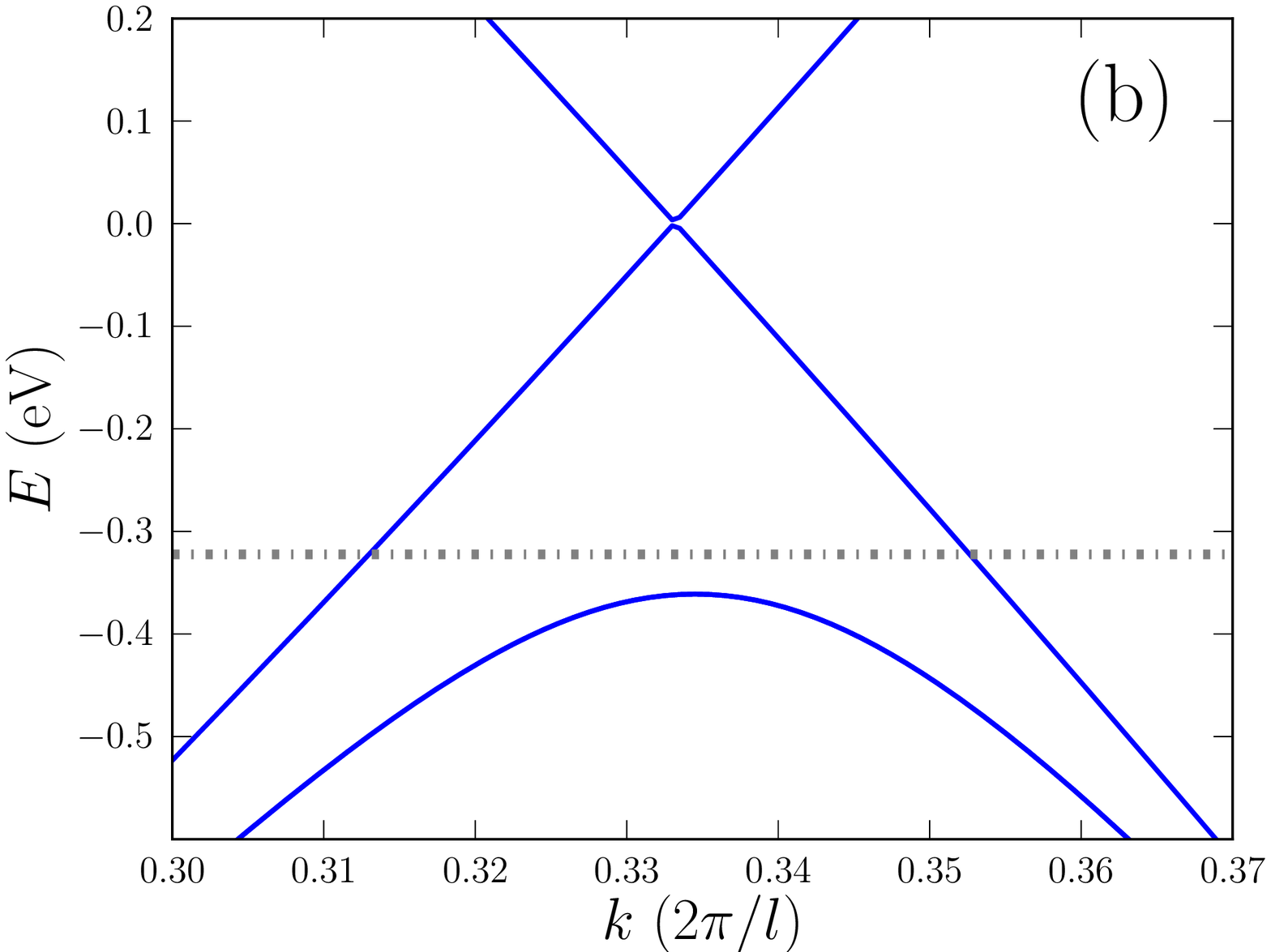}
\includegraphics[width=0.49\columnwidth]{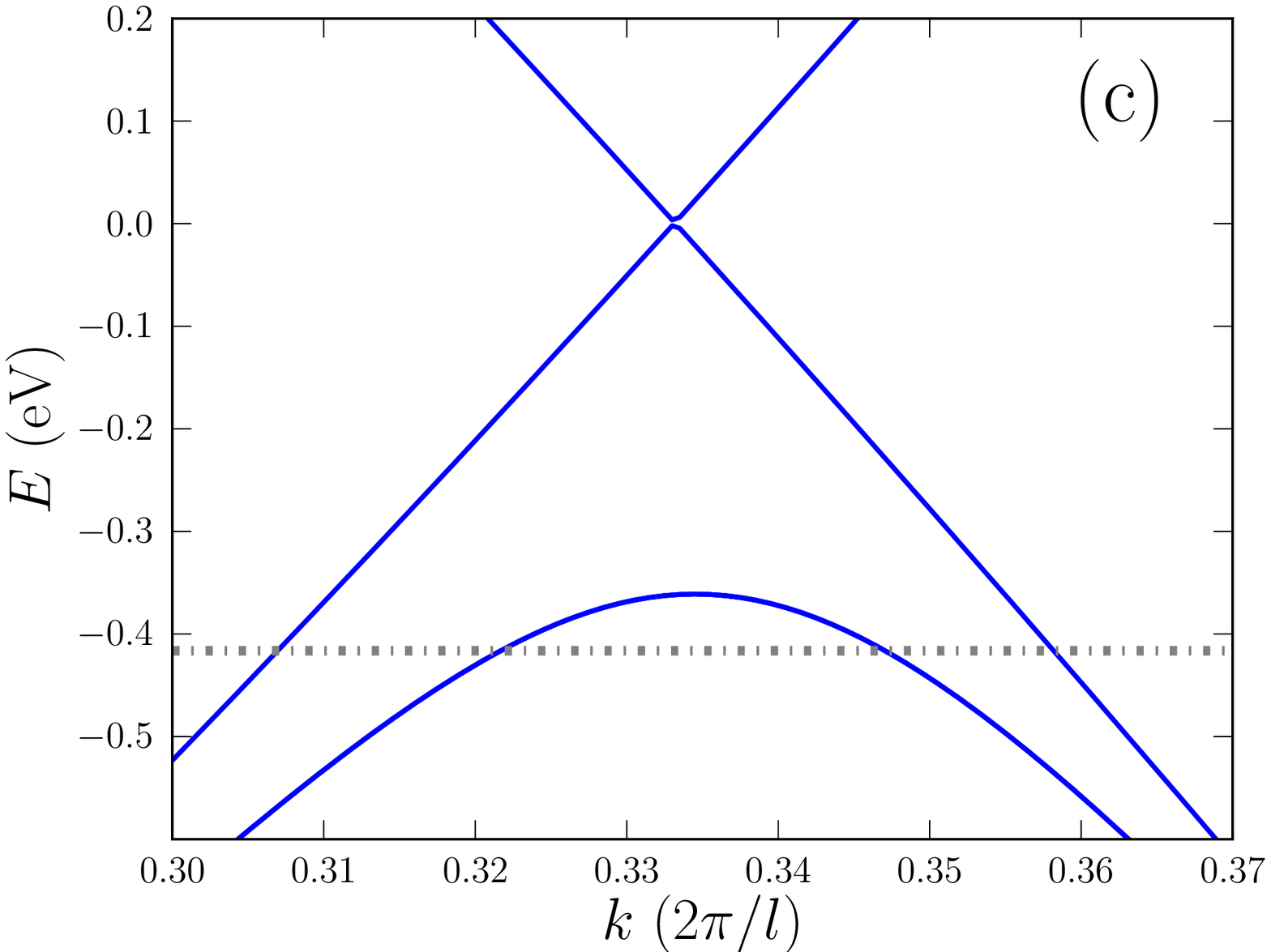}
\includegraphics[width=0.49\columnwidth]{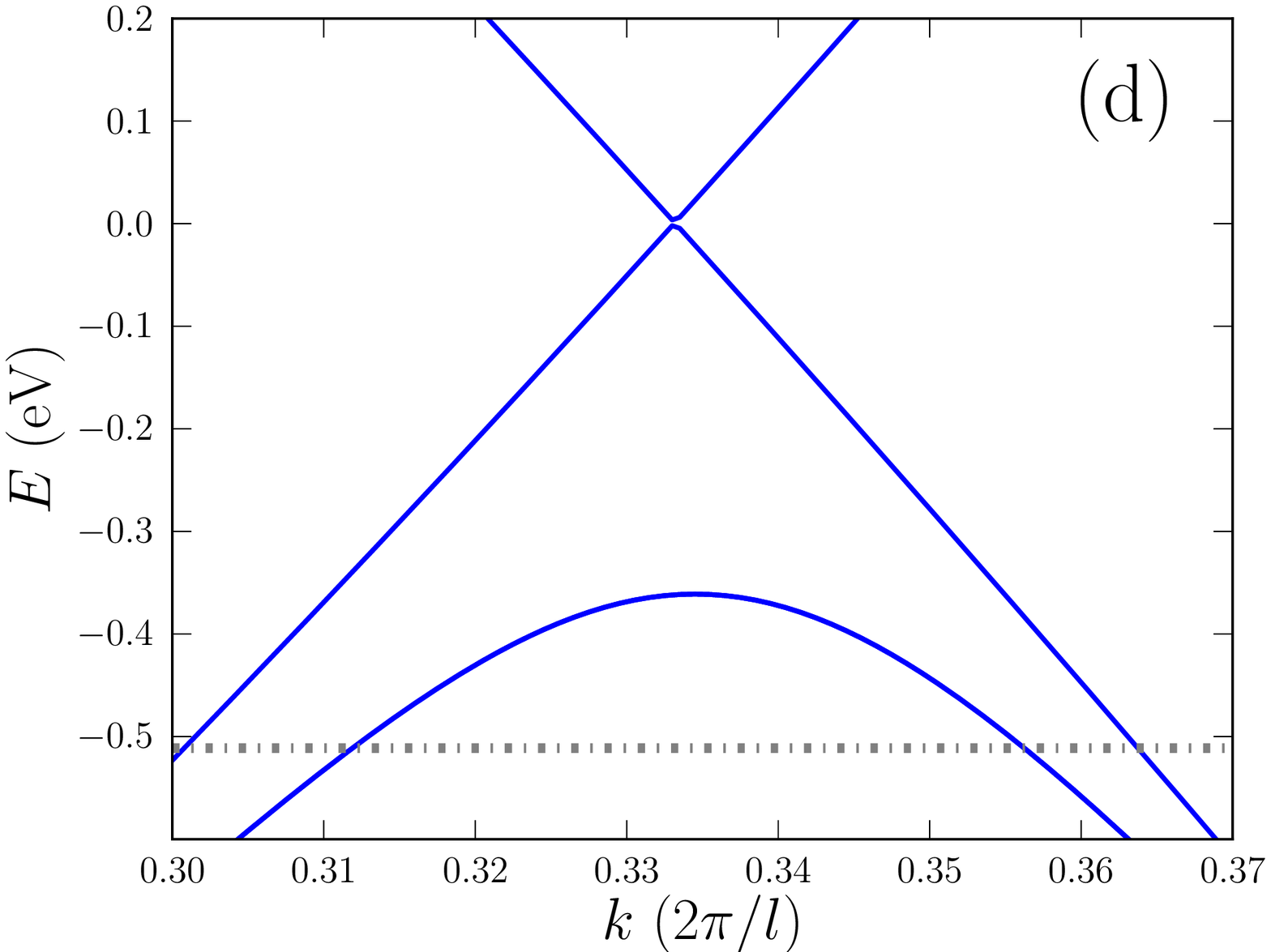}\\
\includegraphics[width=0.49\columnwidth]{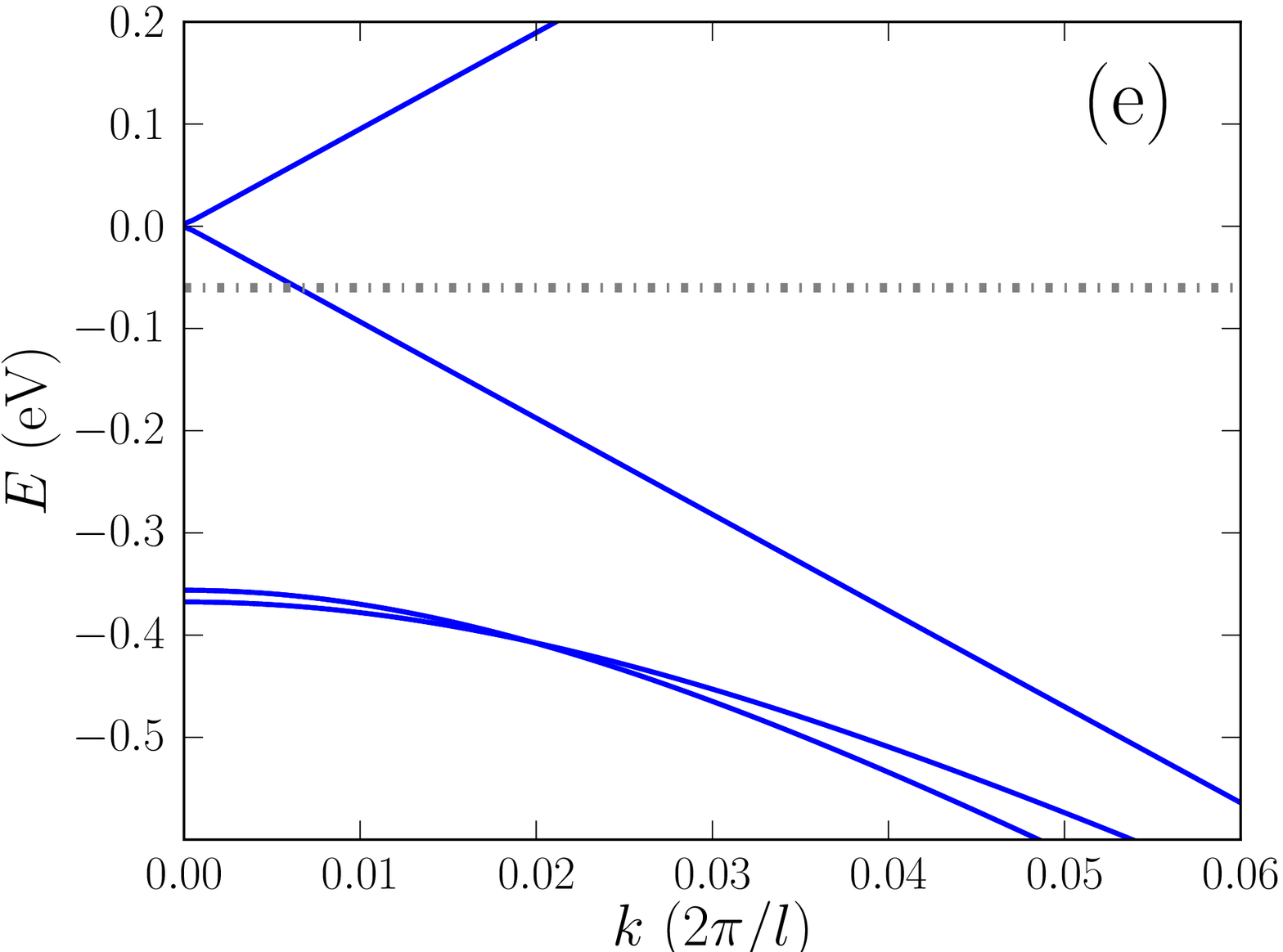}
\includegraphics[width=0.49\columnwidth]{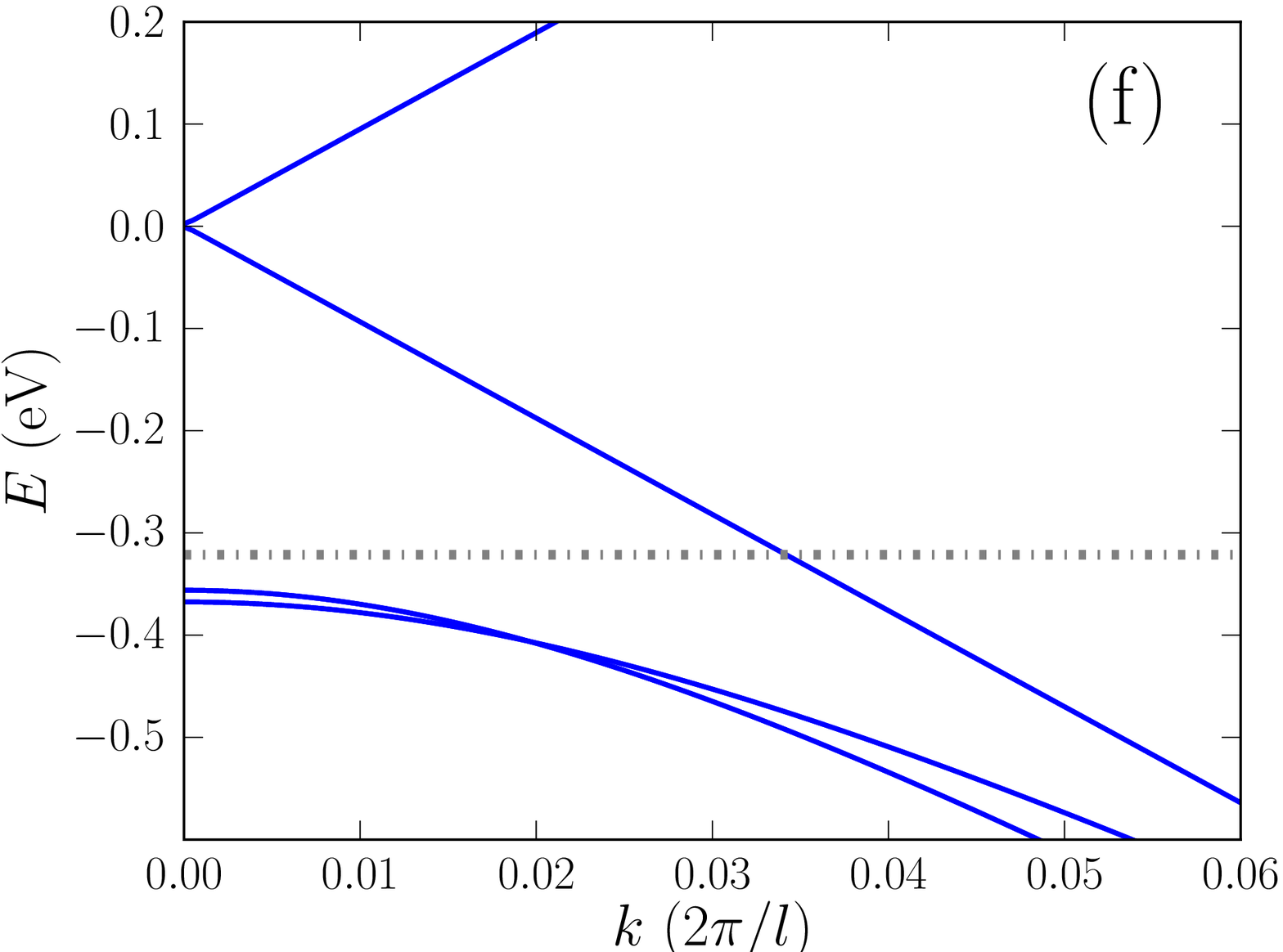}
\includegraphics[width=0.49\columnwidth]{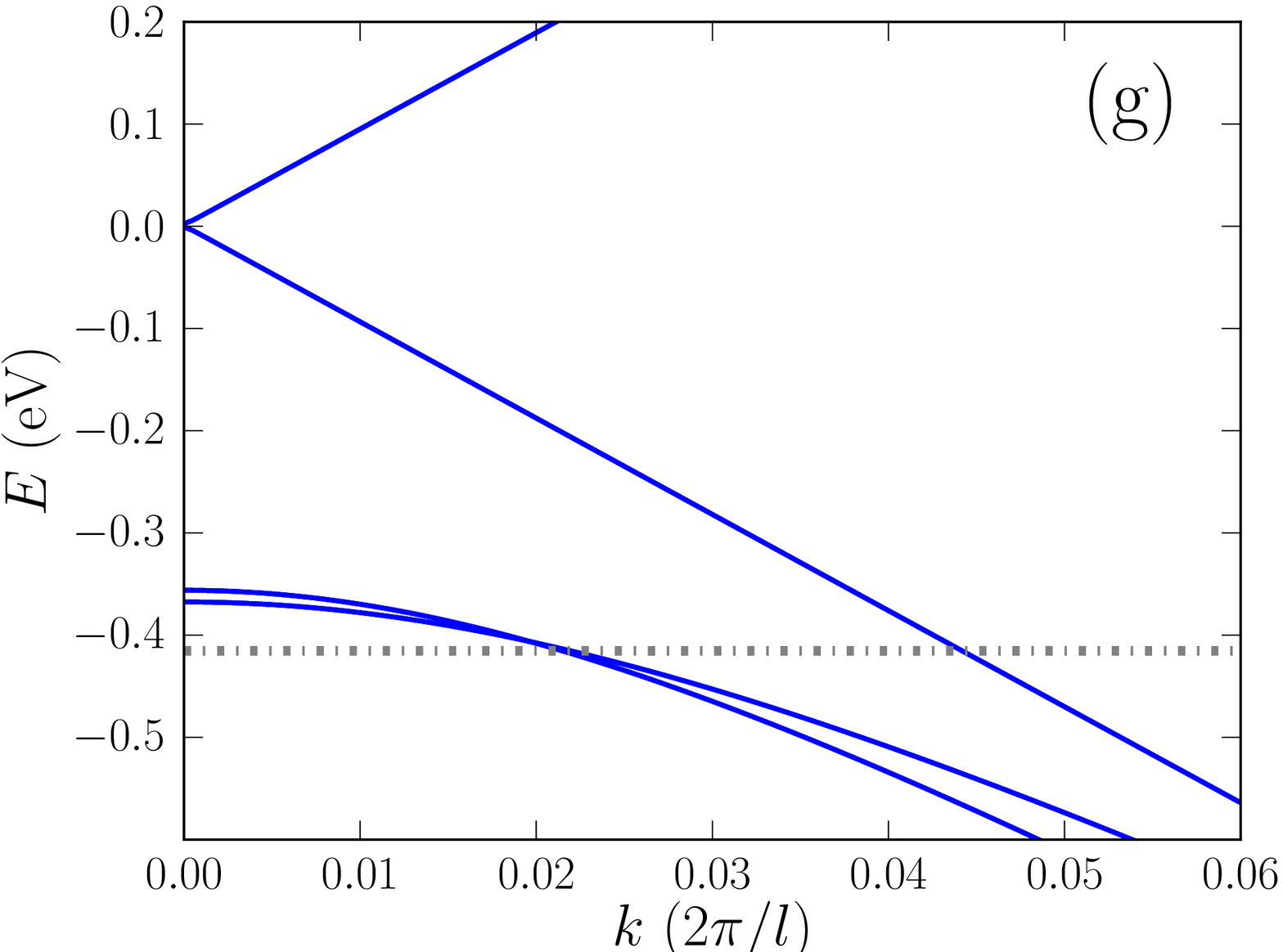}
\includegraphics[width=0.49\columnwidth]{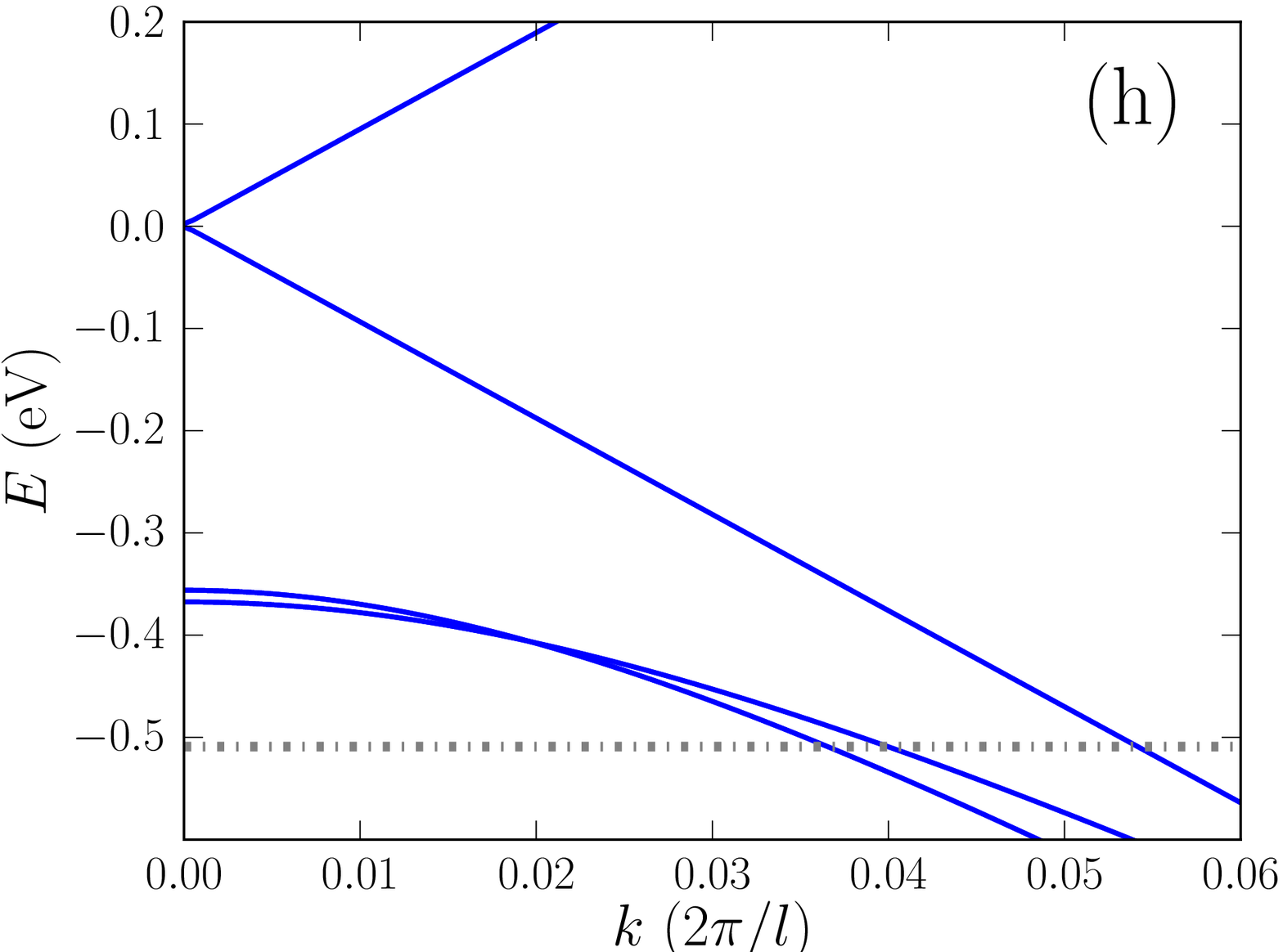}\\
\includegraphics[width=0.49\columnwidth]{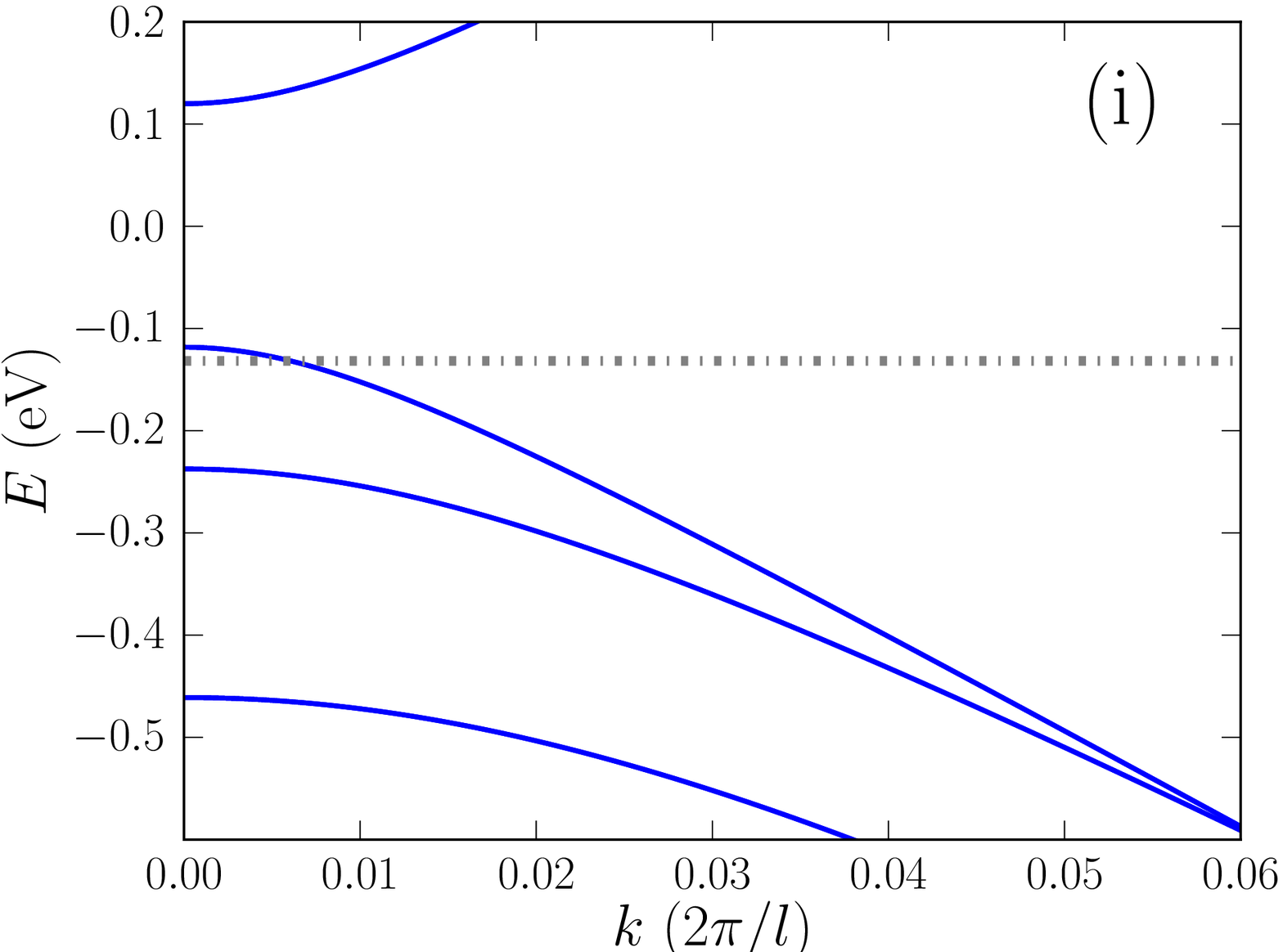}
\includegraphics[width=0.49\columnwidth]{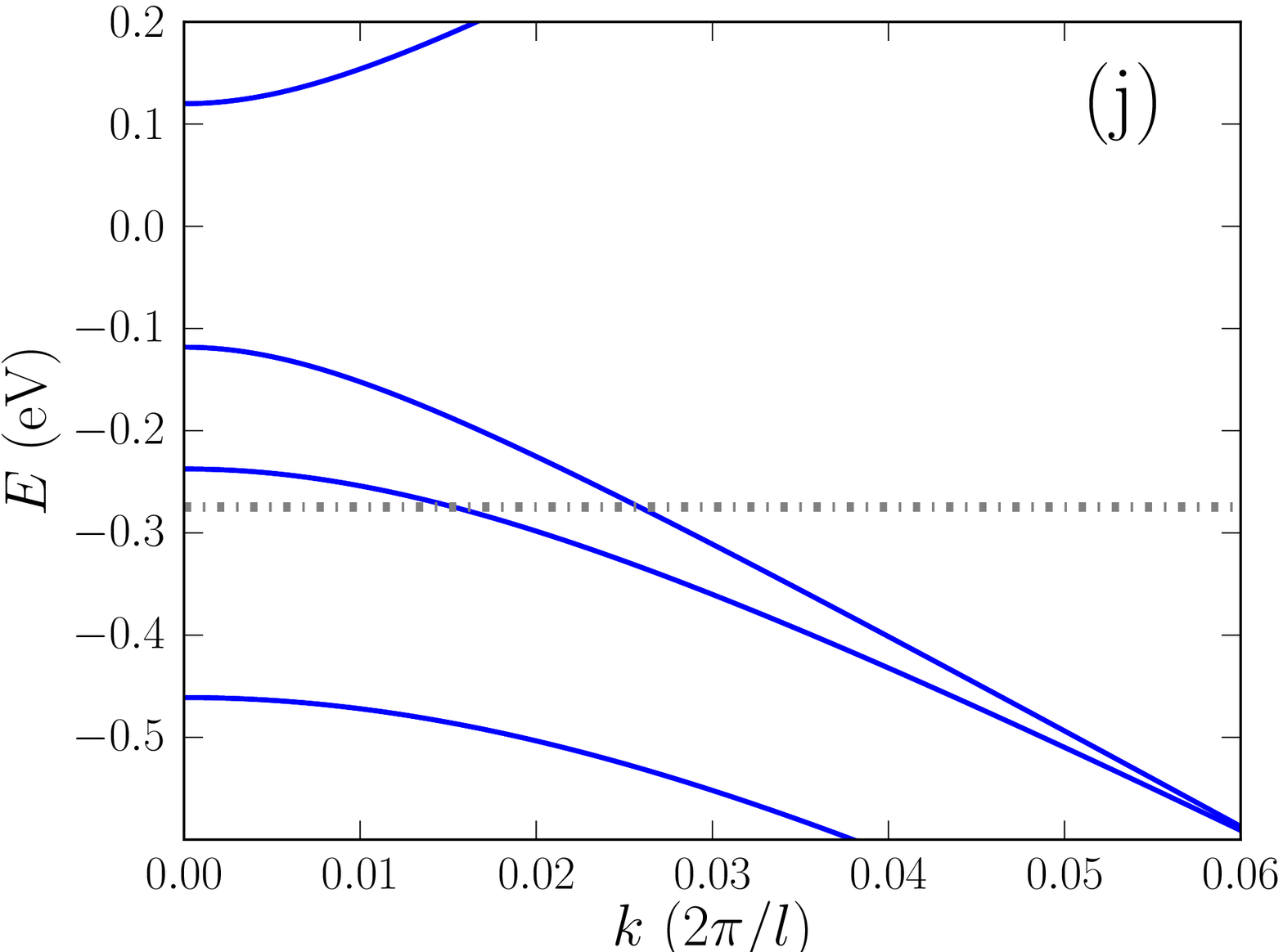}
\includegraphics[width=0.49\columnwidth]{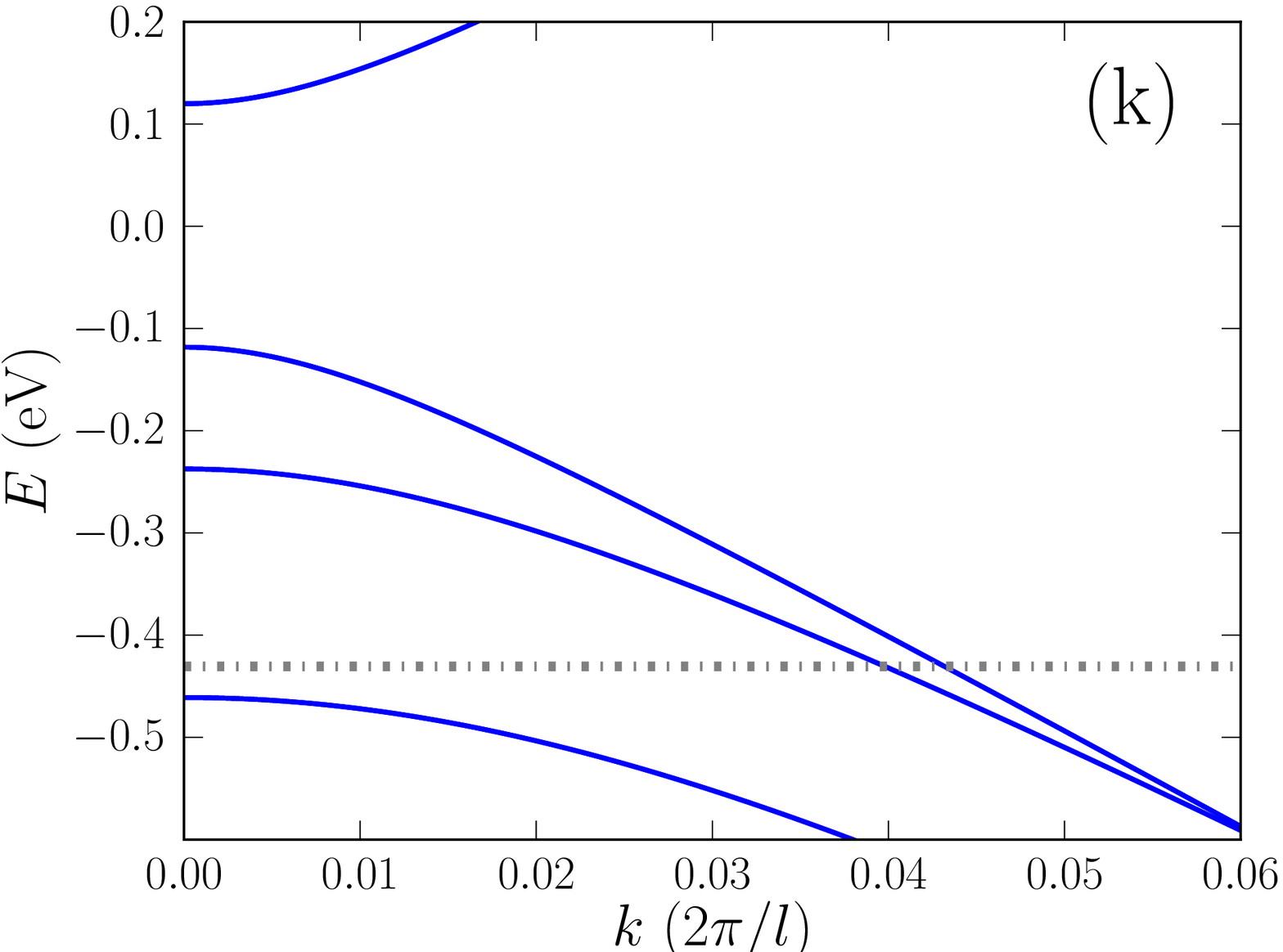}
\includegraphics[width=0.49\columnwidth]{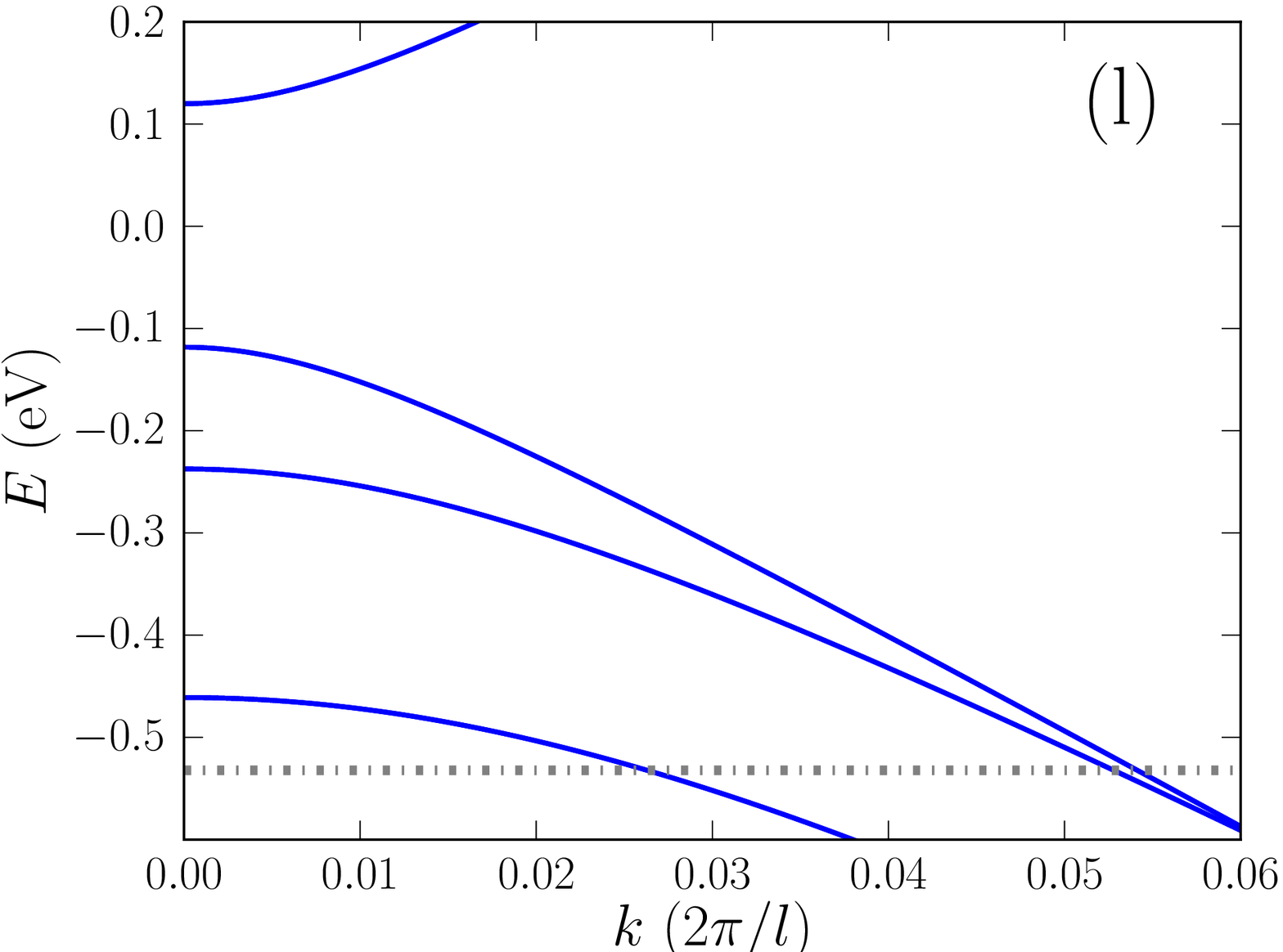}
\caption{Evolution of the Fermi level at $300$~K in the band structure of armchair (a-d), metallic zigzag (e-h) and semiconducting zigzag (i-l) with diameter of $3.5$~nm and 2D effective carrier density of $1$ (a,e,i), $6$ (b,f,j), $13$ (c,g,k) and $20 \times 10^{12}$~cm$^{-2}$ (d,h,l).}
\label{fig_mu_d_300K_change_n}
\end{figure*}

Figure~\ref{fig_mu_d_300K_change_n} shows the position of the Fermi level in the band structure of CNTs with diameter of $3.5$~nm. This figure helps to interpret the results presented in Fig.~\ref{fig_d_n_T}b. At high carrier density, more bands are included in the transport energy window. That not only brings in bands with finite effective mass, but also more scattering mechanisms.

\section{Carrier density dependence of $\mu \times n_{2D}$}
\label{mu_n}

\begin{figure}
\includegraphics[width=0.9\columnwidth]{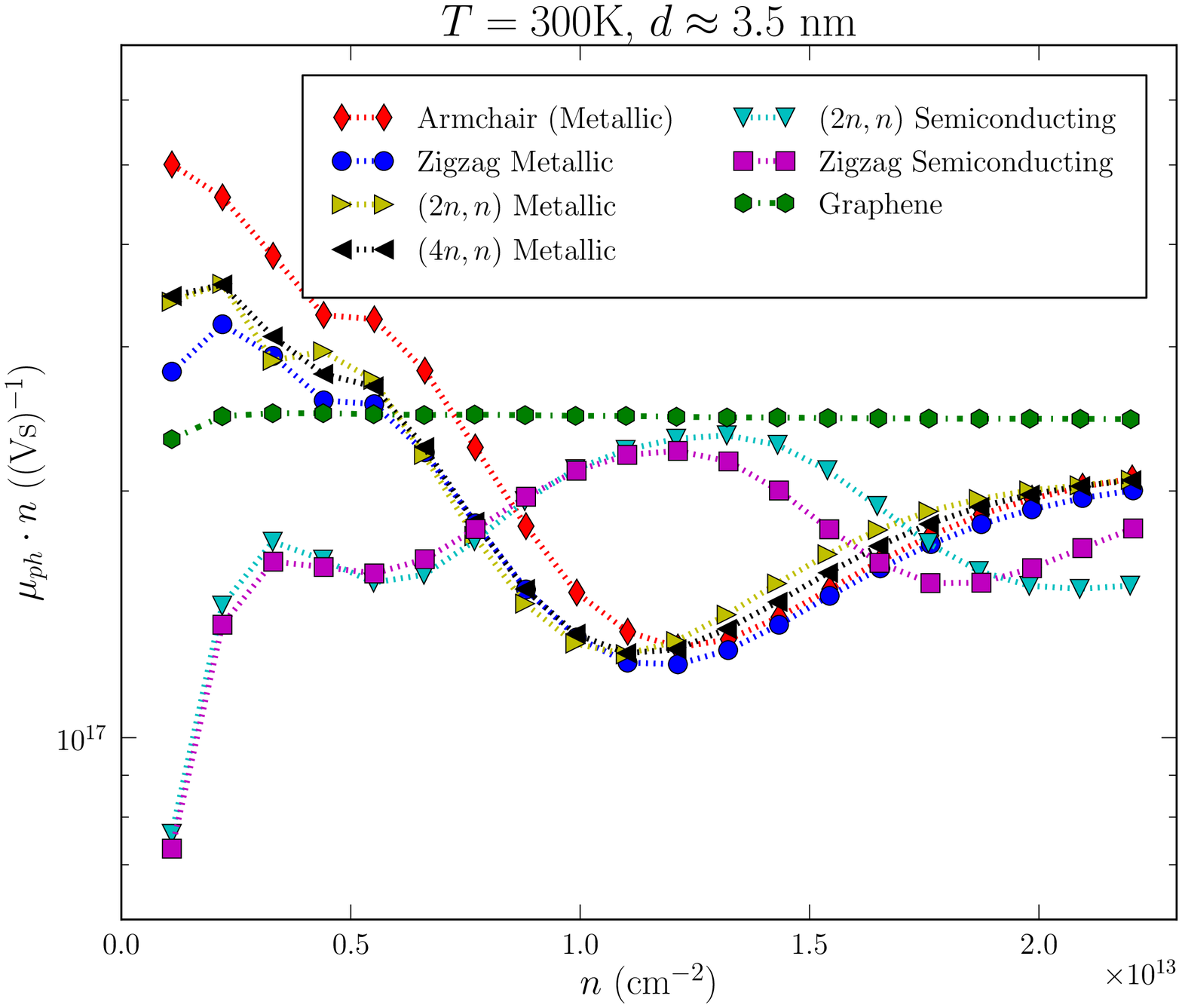}
\caption{Mobility ($\mu$) multiplied by the carrier density ($n_{2D}$) versus $n_{2D}$ in graphene and CNTs ($T=300$~K, $d\approx3.5$~nm). The same data but presented as $\mu$ versus $n_{2D}$ are shown in Fig.~\ref{fig_d_n_T}b.}
\label{fig_mu_n}
\end{figure}

Figure~\ref{fig_mu_n} shows that, at 300~K, the product $\mu \times n_{2D}$ does not depend on the carrier density $n_{2D}$ in graphene. The same behavior is found for CNTs at high carrier density but only approximately, there are deviations coming from band structure effects.

\begin{acknowledgments}
This work was supported by the French National Research Agency (ANR) projects Quasanova (contract ANR-10-NANO-011-02) and Noodles (contract ANR-13-NANO-0009-02). Part of the calculations were run on the TGCC/Curie machine using allocations from GENCI and PRACE. H.M. and L.W. acknowledge support by the National Research Fund (FNR), Luxembourg (projects OTPMD and NanoTMD).
\end{acknowledgments}

\bibliography{graphene_cnt}

\begin{thebibliography}{10}
\bibitem{Note}%
  \BibitemOpen
  \bibinfo {note} {The computation was set up to guarantee an accuracy of at
  least 0.5 meV for the Kohn-Sham eigenvalues. This corresponds to a uniform
  wavelet grid with a spacing of 0.3 bohrs, and a Monkhorst-Pack mesh of 36$^2$
  k-points, for a non-irreductible supercell of 4 carbon atoms. We took care to
  include explicitly the Dirac point in the Brilloiun zone sampling.}%
  \bibAnnoteFile{Stop}{Note}%
\end{thebibliography}

\end{document}